\documentclass[useAMS,usenatbib]{mn2e}
\usepackage{amssymb,amsmath,float,morefloats,graphicx,etex,subfig,array,rotating,lscape,epstopdf}
\DeclareMathOperator*{\argmax}{arg\,max}

\title[Are the Variability Properties of the Kepler AGN Light Curves Consistent with a Damped Random Walk?]{Are the Variability Properties of the Kepler AGN Light Curves Consistent with a Damped Random Walk?}
\author[Vishal P. Kasliwal, Michael S. Vogeley, and Gordon T. Richards]{Vishal P. Kasliwal$^{1}$\thanks{E-mail:
vpk24@drexel.edu (VPK); vogeley@drexel.edu (MSV); gtr@drexel.edu (GTR)}, Michael S. Vogeley$^{1}$\footnotemark[1], \& Gordon T. Richards$^{1}$\footnotemark[1]\\
$^{1}$Department of Physics, Drexel University, 3141 Chestnut St., Philadelphia, PA 19104, USA}

\begin{document}

\date{Accepted 2015 June 1. Received 2015 April 25; in original form 2014 August 1}

\pagerange{\pageref{firstpage}--\pageref{lastpage}} \pubyear{2015}

\maketitle

\label{firstpage}

\begin{abstract}\label{abstract}
We test the consistency of active galactic nuclei (AGN) optical flux variability with the \textit{damped random walk} (DRW) model. Our sample consists of 20 multi-quarter \textit{Kepler} AGN light curves including both Type 1 and 2 Seyferts, radio-loud and -quiet AGN, quasars, and blazars. \textit{Kepler} observations of AGN light curves offer a unique insight into the variability properties of AGN light curves because of the very rapid ($11.6-28.6$ min) and highly uniform rest-frame sampling combined with a photometric precision of $1$ part in $10^{5}$ over a period of 3.5 yr. We categorize the light curves of all 20 objects based on visual similarities and find that the light curves fall into 5 broad categories. We measure the first order structure function of these light curves and model the observed light curve with a general broken power-law PSD characterized by a short-timescale power-law index $\gamma$ and turnover timescale $\tau$. We find that less than half the objects are consistent with a DRW and observe variability on short timescales ($\sim 2$ h). The turnover timescale $\tau$ ranges from $\sim 10-135$ d. Interesting structure function features include pronounced dips on rest-frame timescales ranging from $10-100$ d and varying slopes on different timescales. The range of observed short-timescale PSD slopes and the presence of dip and varying slope features suggests that the DRW model may not be appropriate for all AGN. We conclude that AGN variability is a complex phenomenon that requires a more sophisticated statistical treatment. 
\end{abstract}


\section[]{Introduction}\label{Introduction}

The continuum of the X-ray through optical radiation spectrum observed in active galactic nuclei (AGN) arises in the accretion disk surrounding the central supermassive black hole \citep{sha73a, sha73b}. It is well known that this region of the spectrum exhibits strong, stochastic variability at the $200$ per cent (X-ray) to $10$ per cent (optical) level \citep{mus93, wag95, kro99}. Variability is observed over a wide range of timescales ranging from hours to months to years. The physical mechanisms driving this variability are not well understood --- models range from X-ray flares that drive optical variability \citep{goo06} to local variations in the plasma viscosity of the accretion disk \citep{lyu97} caused by small scale angular momentum outflows triggered by local dynamo processes \citep{kin04}.

The accretion disk of a typical $10^{8} M_{\odot}$ supermassive black hole is too small to be imaged. Models of AGN variability can only be tested by comparing the stochastic properties of observed AGN light curves to the behavior predicted by the variability model. The most popular AGN variability model is the \textit{damped random walk} (DRW) model or 1-parameter Auto-Regressive, AR(1), process \citep*{kel09,koz10,mac10,zu13}. This model is extremely simple and does a good job of fitting ground-based variability data, yet we will demonstrate that it fails to capture the full range of variability behavior exhibited by AGN. We begin by outlining the terminology and mathematics used in discussing variability and the AR(1) process in the next section.   

\section[]{The AR(1) Process}\label{AR1}

The light curve of an AGN may be regarded as a discrete time series that samples a continuous process. Measurements of the source flux $F_{i}$ are made at times $t_{i}$ for $N$ instants in time. Ideally, these measurements are obtained at fixed increments separated by a constant sampling interval $\delta t$. In such cases, the time interval $\Delta t$ between any two flux measurements, $F_{i}$ \& $F_{j}$, is related to the sampling interval by
\begin{equation}\label{eq:Sampling}
\Delta t = n \delta t,
\end{equation} where $n = i - j$ with $1 \leq n < N$. This kind of sampling pattern is usually not possible in the case of ground-based astronomical time series because of interruptions caused by factors such as the weather and the availability of the target in the night sky. Even in the case of space-based measurements of the series, it is possible to have `missing values', i.e. values of $t_{i}$ with $i = n\delta t$, for which there is no corresponding measurement of the flux $F_{i}$. The presence of these missing values complicates the analysis process. In the case of a time series that exhibits stochastic behavior, the goal of time series analysis is to characterize the joint probability distribution of $F_{i}$ for the measured data points. If the joint probability distribution of the data points is independent of time, the time series is said to be \textit{stationary}, i.e. one would observe the same light curve (from a statistical standpoint) at all times.

The timescale on which the bulk fueling rates of the AGN vary is in the hundreds of thousands to millions of years and therefore cannot be directly probed by observation for any individual AGN. It is reasonable to expect that the stochastic variability observed in the flux time series of individual AGN is caused by frequently reoccurring short-lived processes local to the accretion disk, i.e. the time series is stationary on the timescales we probed. It is well known from the theory of time series analysis that a stationary time series can be modeled by an Auto-Regressive Moving Average, or ARMA, process \citep{ham94,woo12,pra10,box06, bro02} of appropriate order (p,q). The general form of an ARMA(p,q) process, i.e. a process with p autoregressive terms and q moving average terms, is given by
\begin{equation}\label{eq:ARpq}
F_{i} = \sum_{m=1}^{p}\phi_{m}F_{i-m} + \sum_{n=1}^{q}\theta_{i-n}w_{i-n} + w_{i},
\end{equation}
where $\phi_{m}$ are autoregression coefficients, $\theta_{n}$ are moving average coefficients, and $w_{i}$ are known as `innovations' in statistical literature. Autoregressive terms introduce infinite-duration correlations in the data while moving average terms introduce finite-duration correlations. It is possible to represent any stationary process as a pure autoregressive or moving average process, albeit perhaps of infinite order \citep{sca81}. Innovations drive the stochastic behavior of the process. Each innovation is an independent random number typically drawn from a Gaussian random distribution with zero mean and variance $\sigma^{2} \propto \delta t$. To make the process stationary we must restrict $| \phi_{i-m} |<1$ for all values of $m$. It is clear from previous work \citep{kel09,koz10,mac10} that AGN light curves exhibit long-term correlations. The simplest ARMA process that exhibits long-term correlations is the purely autoregressive (no moving average terms) AR(1) process or damped random walk (DRW) \citep{ive14} given by 
\begin{equation}\label{eq:AR1}
F_{i} = \phi_{1} F_{i - 1} + w_{i},
\end{equation} 
with $0 < \phi_{1} < 1$. The lone autoregression coefficient $\phi_{1}$ connects the current value of the series to the previous value, making the AR(1) process a \textit{memoryless} or \textit{Markov} process. The numerical value of $\phi_{1}$ depends upon the sampling interval $\delta t$. For a given sampling interval, a value of $\phi_{1} \sim 1$ implies that $F_{i} \simeq F_{i-1}$ and so the value of $\phi_{1}$ sets a characteristic timescale $\tau$ after which the contribution of the previous value of the series will be relatively insignificant, i.e. $\tau$ is the \textit{decorrelation timescale}. Numerically, 
\begin{equation}\label{eq:PhiTau}
\phi_{1} = e^{-\frac{\delta t}{\tau}},
\end{equation}
where $\delta t$ is the sampling interval and $\tau$ is the aforementioned characteristic decorrelation timescale \citep{gil96}. If $\delta t \ggg \tau$, then $\phi \sim 0$ and any given value of the time series is only very weakly correlated with the previous value. On the other hand, if $\delta t \lll \tau$, then $\phi \sim 1$ and the time series exhibits strong correlation. For the series to be stationary, we must have $\phi_{1} < 1$, i.e. $\delta t < \tau$. It turns out that the variance of the innovations is also linked to the decorrelation timescale:
\begin{equation}\label{eq:InnovationVar}
\langle w_{i}^{2} \rangle = \sigma^{2}(1 - e^{-\frac{2 \delta t}{\tau}}),
\end{equation}
where $\sigma$ allows the amplitude of the variance of the innovations to vary independently of $\delta t$ and $\tau$. For a fixed value of $\sigma$, if the sampling interval is much longer than $\tau$, the variance of the innovations will be very close to $\sigma$. Combined with the accompanying small value of $\phi_{1}$, the resulting time series will look close to pure white noise. If, on the other hand, we choose a sampling interval much smaller than the decorrelation timescale, the variance of the innovations is forced to be very close to zero and the value of $\phi_{1}$ is very close to $1$, resulting in very obvious correlations between the data points.

Physically, the AR(1) process could result from an accretion disk with local `spots' that contribute more or less flux than the mean flux level of the disk. These spots appear at random and dissipate over some characteristic physical timescale \citep{ago11}. Under the assumption that the distribution of the contribution to the total flux from each spot is Gaussian, the Central Limit Theorem assures us that the overall changes in flux, i.e. the innovations in Eq. \eqref{eq:ARpq} and in Eq. \eqref{eq:AR1}, are drawn from a Gaussian distribution. Long term correlations exist because the spots do not dissipate instantaneously. The \textit{power spectral density} (PSD) of this time series follows a bent power-law given by
\begin{equation}\label{eq:DRW_PSD}
PSD_{AR(1)}(f) = \frac{4\sigma^{2}\tau}{1 + (2 \pi f \tau)^{2}},
\end{equation}
while the \textit{auto-covariance function} (ACVF), related to the PSD via a Fourier Transform (Wiener-Khintchine theorem), is given by
\begin{equation}\label{eq:DRW_ACF}
ACVF_{AR(1)}(\Delta t) = \sigma^{2}e^{-\frac{\Delta t}{\tau}},
\end{equation}
where $\sigma$ is the low-frequency asymptotic amplitude and $\tau$ is the characteristic timescale at which the PSD turns over and decays \citep{gil96}. Hence the \textit{auto-correlation function} (ACF) is given by
\begin{equation}\label{eq:DRW_ACF2}
ACF_{AR(1)}(\Delta t) = e^{-\frac{\Delta t}{\tau}},
\end{equation}

Does the AR(1) process actually describe the variability of AGN? Recent ground based studies \citep{zu13,gra14} indicate that on very short timescales, AGN light curves show strong autocorrelation with PSD slopes steeper than $1/f^{2}$. To model stochastic light curves with arbitrary PSD slopes, we introduce the \textit{damped power-law} (DPL) model by generalizing the AR(1) PSD in Eq. \eqref{eq:DRW_PSD} to 
\begin{equation}\label{eq:DPL_PSD}
PSD_{DPL}(f) = \frac{\sigma_{Eff}^{2}}{1 + (2 \pi f \tau)^{\gamma}},
\end{equation}
where we have lumped the quantities in the numerator of Eq. \eqref{eq:DRW_PSD} into a single variable $\sigma_{Eff}$ and introduced $\gamma$ to allow the logarithmic slope of the PSD to be a free parameter. When $\gamma < 2$, the process exhibits weaker autocorrelation on short timescales than the AR(1) i.e. the time series looks less smooth. When $\gamma > 2$, the process shows stronger autocorrelation on short timescales than the AR(1) i.e. the time series looks smoother. This DPL model is a simplification of the popular `Bending Power-Law' (BPL) model of \cite{mch04}:
\begin{equation}\label{eq:McHardyPSD}
PSD_{BPL}(f) = Af^{-\alpha_{1}}\prod_{i=1}^{L}\frac{1}{1+(\frac{f}{f_{Bi}})^{\alpha_{i+1}-\alpha_{i}}},
\end{equation}
where $\alpha_{i}$ are the logarithmic PSD slopes above the corresponding `bend' frequencies $f_{Bi}$. Allowing the logarithmic PSD slope to be a free parameter is similar to the structure function parametrization chosen by \cite{sch10}:
\begin{equation}\label{eq:SchmidtSF}
SF(\Delta t) = A \left( \frac{\Delta t}{1 \ yr} \right)^{\lambda},
\end{equation} where $\lambda$ plays the same role as $\gamma$ in the DPL model--they allow the correlation level at fixed time-lag to vary from that predicted by a pure AR(1) process. To test our DPL model, we use light curve data from the NASA \textit{Kepler} mission.

\section[]{The \textit{Kepler} Mission}\label{KeplerMission}

The \textit{Kepler} mission was designed to survey the Miky Way for Earth-sized and smaller planets around the habitable zone by using the transit-method to detect exo-planets. \textit{Kepler} is designed as a Schmidt-camera with a primary aperture of 0.95-m, a fixed 115.6 deg$^{2}$ FOV, and a half-maximum bandpass from 435-nm to 845-nm with $\ge$ 1 per cent relative spectral response from 420 nm to 905 nm \citep{KIH}. The \textit{Kepler} focal-plane consists of 42 CCDs arranged in 21 modules that are readout by 4 channels each. Each $50 \times 25$-mm CCD has $2200 \times 1024$ pixels yielding an image scale of 3.98 arcsec per pixel. The \textit{Kepler} PSF is intentionally large--6.4 pixels near the center of the FOV--in order to increase the photometric precision \citep{gil04} by ensuring that no pixel contains more than 60 per cent of the flux from a point source target. This has two consequences that complicate simple aperture photometry: 1. point sources, such as stars, appear extended, and 2. faint background sources `crowd' the aperture used to measure the flux from a target of interest. The signal at each pixel on the CCDs is measured over an interval known as the \textit{integration time} or \textit{frame} which consists of a variable \textit{exposure time} and a fixed \textit{readout time} of 0.520 s. The flight exposure time is 6.02 s yielding a flight integration time of 6.54 s per frame. Two datasets can be created by \textit{Kepler}: the \textit{long cadence (LC) data} and the \textit{short cadence (SC) data}. The long cadence data is integrated over 270 frames (29.430 min), while the short cadence data is integrated over 9 frames (0.981 min). Most of the objects observed by us have no SC data--only LC data was obtained. Furthermore, the SC data set is harder to calibrate because of the small number of SC quiet targets observed \citep{kin12}. For these reasons we only use LC data in our analysis. \cite{kin12} provides an excellent overview of the final data products supplied by \textit{Kepler}, including strategies for dealing with the various types of artifacts present in the data. 

The LC data-set poses several calibration challenges. Certain readout channels on specific CCD modules have known performance issues discussed in the \cite{KIH} including out-of-spec read noise levels and gain. More troubling are the `moire' and `rolling band' effects seen in some channels \citep{kol10}. The combination of module and channel is usually quoted in the format module\#.channel\#. A \textit{Kepler} target will fall on a fixed sequence of 4 module and channel combinations over the course of a year, i.e. since the spacecraft rolls 90$^{o}$ every quarter, at the beginning of every 5$^{th}$ quarter targets will fall on the same module\#.channel\# combination and will be subject to the same instrumentation issues. We discuss the impact of these effects in Sec. \ref{StitchResults}.

Spacecraft operation events and systematic trends such as thermally-driven focus variations, pointing offsets, and differential velocity aberration (DVA), introduce artifacts into the data; further systematics are introduced as the result of light losses caused by the time dependent sampling of the outer wings of the target PSF and contamination from neighboring sources \citep{KDCH,KIH}. Such artifacts can be misinterpreted as being astrophysical in origin, and can mask true astrophysical signals. Post-calibration \textit{Kepler} data is available in both uncorrected (SAP flux) and corrected (PDCSAP flux) forms. Corrections to the calibrated data are obtained using the PDC module \citep{stu12,smi12}.

The PDC module identifies features common to hundreds of quiet targets on each CCD by examining the cross-correlation between targets. The most correlated targets are used to create 16 best-fit vectors called ÒCotrending Basis VectorsÓ (CBVs) using Singular Value Decomposition (SVD). Corrections to the calibrated data are obtained by removing the CBVs from the data using a weighted normalization. The algorithm used by PDC to compute these CBV weights  changed with the introduction of Ver 8.0 of the \textit{Kepler} Science Operations Center (SOC) pipeline (Data Release 14). Prior to pipeline Ver. 8.0, the PDC module used a Least Squares (PDC-LS) approach to compute the CBV weights. Ver. 8.0 introduced a Bayesian \textit{Maximum A Posteriori} (PDC-MAP) algorithm compute the CBV weights and optimally remove spacecraft-induced trends while minimizing the removal of the true underlying signal \citep{stu12}. The Bayesian-MAP procedure assumes that the the observed signal may be represented as 
\begin{equation}\label{eq:BayesianMAPModel}
F = H \theta + \epsilon,
\end{equation}
where $H$ is a design matrix consisting of the CBV vectors (typically 4), $\theta$ is the vector of the CBV weights (which is what the PDC module is trying to determine), and $\epsilon \sim (0,\Sigma_{N})$ is the (uncorrelated, i.i.d.) instrument noise vector. The log-likelihood of an observed light curve, assuming this model for the data, is then given by
\begin{equation}\label{eq:LCLnLikelihood}
\ln p(F \vert \theta) = -\frac{N \ln(2 \pi) + \ln(\vert \Sigma_{N} \vert) + (F - H \theta)^{T} \Sigma_{N}^{-1} (F - H \theta)}{2}.
\end{equation}
Maximization of this likelihood can be performed analytically, however the resulting de-trended light curve \textit{will necessarily} have intrinsic variability removed because LS calculates the CBV weights by projecting the raw light curve onto each CBV---any coincidences between the true light curve and the CBV will spuriously add to the CBV weight.

To avoid this behavior, the new PDC-MAP approach applies a weighted Bayesian prior, $p(\theta)$, to the likelihood in \eqref{eq:LCLnLikelihood} to bias the best-fit CBV weights toward removing only the spacecraft-induced variability signal while leaving intrinsic variability intact. $p(\theta)$, is constructed by using LS to de-trend quiet targets in the phase-space vicinity (RA, Dec, and apparent magnitude) of the object of interest. The prior weight, $W_{pr}$, is calculated such that it is closer to 0 for relatively quieter targets and closer to 1 for relatively active targets. The PDC algorithm computes
\begin{equation}\label{eq:PDCLikelihood}
\widehat{\theta} = \argmax_{\theta} (\ln p(F \vert \theta) + W_{pr} \ln{p(\theta)}).
\end{equation}
Therefore $p(\theta)$ biases the CBV weights towards values corresponding to those for neighboring quiet targets. Since the CBV weights are biased towards removing just enough features in the light curve to render featureless light curves for neighboring quiet targets, intrinsic variability in the light curve of a truly variable object should be relatively unaffected by the de-trending procedure. Further details on the PDC de-trending algorithm may be found in \cite{smi12}.

The benefit of using the \textit{Kepler}-MAST supplied PDCSAP flux light curves is that for most targets, the PDC-MAP algorithm produces optimally de-trended light curves free of instrument effects. At the same time, the drawback of doing so is that it is theoretically possible for the PDC-MAP pipeline to remove intrinsic variability, particularly when the target exhibits intrinsic long-term trends that cancel out the systematic trend, or when the location of the target in position-luminosity phase-space is lacking an adequate number of close neighbors. On the other hand, using the raw SAP flux light curves and performing a customized de-trending using the PyKE tool \textit{kepcotrend} implies that the CBV weights must be computed using no information from neighboring targets and hence are likely to project out intrinsic variability in the light curve.  

We use \textit{Kepler}-MAST supplied light-curves from Data Release 23 that have been processed using version 9.1 of the \textit{Kepler} SOC pipeline (PDC module uses PDC-MAP algorithm) and only perform a minor correction to remove the large quarterly offsets in the data caused by spacecraft rolls.

\section[]{Sample Selection}\label{SampleSelection}

Only $\sim 7$ AGN were known to lie in the \textit{Kepler} FOV prior to the start of the mission due to the lack of coverage from existing extragalactic surveys. Since the beginning of the mission in May 2009, Guest Observer (GO) efforts to find more AGN in the \textit{Kepler} FOV by multiple groups \citep{car12,ede12,weh13} have led to a total of roughly 80 AGN of various types being monitored by \textit{Kepler}. The May 2013 failure of a critical reaction wheel led to the termination of the original \textit{Kepler} mission. We wish to examine variability properties across a wide range of AGN type and redshift; however a large number of the \textit{Kepler} AGN were selected photometrically and do not have redshifts. As such, we restrict our analysis to the 20 that are spectroscopically confirmed AGN. This allows us to reject non-AGN contaminants, and also allows us to study variability behavior as a function of AGN type and perform comparisons in the rest-frame of the object. 

\begin{table*}
\setlength{\tabcolsep}{4pt}
\caption{\label{tab:TargetList} \textit{Kepler} AGN Sample.}
\begin{tabular}{@{}rrcclccc@{}}
\hline
\hline
\multicolumn{2}{c}{Identifier} & \multicolumn{5}{c}{Physical Parameters} \\
KeplerID & Alt. ID & RA & Dec & z & $M_{g}$ & AGN Type \\
 &  & Hrs min sec & Deg min sec & & & \\
\hline
  6932990 &                  Zw229-15$^{\dagger}$ & 19 05 25.969 & $+$42 27 40.07 & 0.0275 & -23.8                        & Sy1$^{\dagger}$ \\
12158940 &                          1925+50$^{\star}$ & 19 25 02.181 & $+$50 43 13.95 & 0.067   & -21.9                        & Sy1 \\
11178007 &                W2R 1858+48$^{\star}$ & 18 58 01.111 & $+$48 50 23.39 & 0.079   & -20.2                        & Sy1 \\
\hline
   9650715 & RXS J19298+4622$^{\dagger}$ & 19 29 50.490 & $+$46 22 23.59 & 0.127   & -21.8$^{\dagger}$ & Sy1$^{\dagger}$ \\
   2837332 &                W2R 1910+38$^{\star}$ & 19 10 02.496 & $+$38 00 09.47 & 0.130   & -20.4                        & Sy1 \\
   3347632 &                W2R 1931+38$^{\star}$ & 19 31 15.485 & $+$38 28 17.29 & 0.158   & -21.0                        & Sy2$^{?}$ \\
   5781475 &                W2R 1915+41$^{\star}$ & 19 15 09.127 & $+$41 02 39.08 & 0.222   & -22.1                        & Sy1 \\
\hline
   2694186 &                W2R 1904+37$^{\star}$ & 19 04 58.674 & $+$37 55 41.09 & 0.089   & -23.8                        & Sy1$^{?}$ \\
   9215110 &                   W2 1922+45$^{\star}$ & 19 22 11.234 & $+$45 38 06.16 & 0.115   & -22.2                        & Sy1.9$^{\star}$ \\
 10841941 &                W2R 1845+48$^{\star}$ & 18 45 59.577 & $+$48 16 47.57 & 0.152   & -21.2                        & Sy1$^{?}$ \\ 
   3337670 &                W2R 1920+38$^{\star}$ & 19 20 47.750 & $+$38 26 41.28 & 0.368   & -23.0                        & Sy1 \\ 
   7610713 &                          1931+43$^{\star}$ & 19 31 12.566 & $+$43 13 27.62 & 0.439   & -24.7                        & QSO \\
\hline
   6690887 &                W2R 1926+42$^{\star}$ & 19 26 31.089 & $+$42 09 59.12 & 0.154   & -21.7                        & BL-Lac$^{\star,\S}$ \\
   6595745 &                W2R 1914+42$^{\star}$ & 19 14 15.492 & $+$42 04 59.88 & 0.502   & -24.6                        & QSO \\
   5597763 &                W2R 1853+40$^{\star}$ & 18 53 19.284 & $+$40 53 36.42 & 0.625   & -25.1                        & QSO \\
 11606854 &         MG4 J191843+4937$^{\P}$ & 19 18 45.617 & $+$49 37 55.06 & 0.926   & -25.0                        & FSRQ$^{\ddagger}$ \\
 10663134 &           Q 1922+4748$^{\dagger}$ & 19 23 27.234 & $+$47 54 17.00 & 1.520   & -25.2                        & QSO$^{\dagger}$; FSRQ$^{\ddagger}$; Jet$^{\|}$ \\
\hline
   8703536 &   IGR J19473+4452$^{\dagger}$ & 19 47 19.308 & $+$44 49 42.32 & 0.0539 & -21.7                       & Sy2$^{\dagger}$ \\
 11021406 &         6C 1908+4829$^{\dagger}$ & 19 09 46.501 & $+$48 34 32.26 & 0.513   & -23.2                       & Sy1.5$^{\dagger}$; FSRQ$^{\ddagger}$ \\
 12208602 &                    4C 50.47$^{\dagger}$ & 19 26 06.318 & $+$50 52 57.14 & 1.098   & -24.7                       & QSO$^{\dagger}$; Jet$^{\|}$ \\ 
\hline
\end{tabular}

$^{\star}$Reliable Identifications of Active Galactic Nuclei from the \textit{WISE}, 2MASS, and \textit{ROSAT} All-Sky Surveys \citep{ede12}

$^{\dagger}$A Catalog of Quasars and Active Nuclei: 13th Ed. \citep{vcv10}

$^{\ddagger}$An All-Sky Survey of Flat-Spectrum Radio Sources \citep{hea07}

$^{\S}$Kepler Observations of Rapid Optical Variability in the BL Lac Object W2R1926+42 \citep{ede13}

$^{\|}$A New List of Extra-Galactic Radio Jets \citep{liu02}

$^{\P}$CGRaBS: An All-Sky Survey of Gamma-Ray Blazar Candidates \citep{hea08}

$^{?}$Weak spectral features--unsure sub-classification

\medskip

The AGN are grouped based on the light curve categories discussed in Sec. \ref{StitchResults}. Col. 6 lists SDSS g-band absolute magnitudes computed from the SDSS apparent magnitude supplied in the \cite{bro11} with cosmological parameters $H_{0} = 67.77 \ \mathrm{km} \mathrm{s}^{-1} \mathrm{Mpc}^{-1}$, $\Omega_{M} = 0.3071$, and $\Omega_{\Lambda} = 0.6914$. We performed a \textit{k-correction} using Eq. (10.29) in \cite{pet03} assuming $\alpha = -0.5$ \citep{ric06} to compute the absolute magnitude at $z = 0$. We present tentative sub-classifications for 11 AGN in Col. 7 using imaging from the STScI Digitized Sky Survey, SDSS and spectra in \cite{ede12}. In the case of broad-line objects we distinguish between Seyfert 1s and QSOs based on the presence of a galactic component in STscI DSS/SDSS imaging and also based on $M_{g} > -24$ for Seyferts.
\end{table*}

Of the 7 AGN known to lie in the \textit{Kepler} FOV prior to 2009, the best studied is kplr006932990, also known as Zw 229-15. The VCV Catalog \citep{vcv10} lists kplr006932990 as a radio-quiet Type 1 Seyfert at a redshift of 0.027 with V-band absolute magnitude $M_{V} = -19.9$. Based on H$\beta$ reverberation mapping, \cite{bar11} obtain a virial black hole mass of $M_{BH} = 1.00^{+0.19}_{-0.24} \times 10^{7} M_{\odot}$. The other Type 1 AGN known to exist in the \textit{Kepler} FOV prior to the commencement of the mission is the X-Ray source kplr09650715 (RXS J19298+4622), which is a radio-quiet Seyfert 1 at $z = 0.127$ with V-band absolute magnitude $M_{V} = -21.8$ \citep{vcv10}. Two narrow-line objects were known to exist in the \textit{Kepler} FOV: kplr008703536 (IGR J19473+4452), which is a radio-quiet Seyfert 2 \citep{mas06} at $z = 0.0539$ with V-band absolute magnitude $M_{V} = -21.1$ \citep{vcv10}, and kplr011021406 (6C 1908+4829), which is a radio-loud object at $z=0.513$ with photographic magnitude $M_{O} = -21.9$ variously classified as either a Seyfert 1.5 by \cite{vcv10} or a FSRQ by \cite{hea07}. At higher redshifts, pre-\textit{Kepler} AGN include the FSRQ blazars kplr11606854 (MG4 J191843+4937) at $z = 0.926$ and kplr010663134 (Q 1922+4748) at $z = 1.52$ \cite{hea07}. The last of the pre-\textit{Kepler} AGN is the quasar kplr012208602 (4C 50.47) at a redshift of $1.098$. Spectra obtained using the MMT indicates that this object has an exceptionally strong CIV 1549 line with a rest-frame equivalent width of 240 \AA \ and a power-law spectrum from 178 MHz to 5 GHz with spectral index $\alpha = -0.65$ \citep{wal84}. Both kplr012208602 and kplr010663134 possess radio-jets \citep{liu02}.

\cite{ede12} obtained spectra of the remaining 13 AGN in our sample using the Kast double spectrograph on the Lick 3 m telescope. \cite{ede12} classify kplr006690887 as a BL Lac object based on a featureless spectrum and coincidence with the radio source NVSS J192631+420958, and classify kplr09215110 as a Type 1.9 Seyfert. Since no sub-classification is provided for the remaining 11 objects, we performed a visual inspection of the spectra provided by \cite{ede12} to characterize these objects by emission line type. We conclude that while most of the remaining objects are Type 1 AGN with prominent broad emission lines, kplr003347632 may be a Type 2 object. Although no SDSS spectra exist for the AGN in our sample, 4 of the new AGN (kplr002694186, kplr002837332, kplr003337670, and kplr005597763) land within the SDSS DR10 footprint. By combining SDSS with STScI DSS imaging, we sub-classify the objects as Seyferts or QSOs based on visual appearance and k-corrected absolute magnitude (traditionally $M_{i} \lesssim -22$ for QSOs from \cite{ric06}) computed from the \cite{bro11} SDSS g-band apparent magnitude available on MAST.

Table \ref{tab:TargetList} presents positional and physical parameters for the light curves from Fig. \ref{fig:AllLightcurves}. Column 1 lists the \textit{KeplerID} used when querying the \textit{Kepler} MAST database \citep{KAM}. Column 2 lists the `common' name of the object along with a primary reference for the object, usually drawn from either \cite{vcv10} or \cite{ede12}. The next two columns supply the J2000.0 RA and Dec location of the object as listed in the \cite{bro11}. The next column supplies the redshift of the object as listed in the primary reference for the object. The next column supplies the \textit{k-corrected} SDSS g-band absolute magnitude ($M_{g}$) based on the apparent SDSS g-band KIC magnitude ($m_{g}$). As described in the \cite{bro11}, the galactic extinction corrected $m_{g}$ was determined during the Stellar Classification Project (SCP) by using DAOPHOT \citep{ste87} to perform PSF photometry on star-like objects. We have converted these g-band magnitudes to absolute magnitudes at $z = 0$ using \textit{k-corrections} from \cite{pet03} with cosmological parameters $H_{0} = 67.77 \ \mathrm{km} \mathrm{s}^{-1} \mathrm{Mpc}^{-1}$, $\Omega_{M} = 0.3071$, and $\Omega_{\Lambda} = 0.6914$, and assuming $\alpha = -0.5$ \citep{ric06}. Column 7 lists the AGN type of the object either from the primary reference or based on a visual examination of imaging and spectra from publicly archived data \citep{bro11,ede12}.  

Light curves from \textit{Kepler} exhibit huge quarterly offsets and require `stitching' before we may analyze them. The next section discusses how we stitch the light curves together.
 
\section[]{Stitching Light Curves Across Quarters}\label{Stitching}

\begin{figure}
\center{\includegraphics[width=\columnwidth]{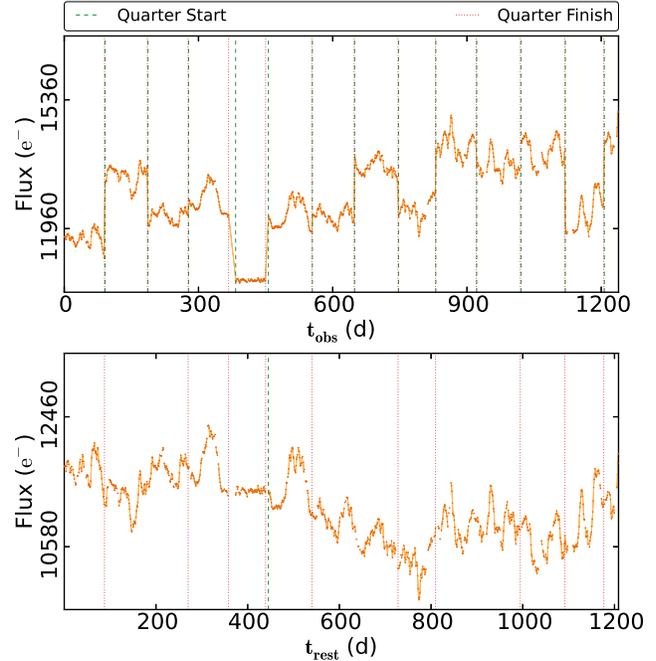}}
\caption{\label{fig:kplr006932990-Lightcurves} PDCSAP light curve of the Seyfert 1 kplr006932990 (Zw 229-15) --- Top panel: unstitched; bottom panel: stitched. The huge discontinuities in flux in the top panel are caused by redefinition of the target aperture after the spacecraft performs a roll maneuver as explained in Sec. \ref{Stitching} and are endemic to all the \textit{Kepler} light curves. Our stitching technique serves to remove the quarterly discontinuities present in the light curve obtained from MAST. }
\end{figure}

As discussed in Sec. \ref{KeplerMission} and \ref{SampleSelection}, quarterly discontinuities exist in \textit{Kepler} data. Once every 3 months, the spacecraft performs a roll maneuver to transmit data to the Earth. After each roll maneuver, targets fall on different CCDs resulting in slightly different distributions of the target flux across adjacent pixels. To compensate, the target aperture is redefined after every roll maneuver. Due to the large size of the \textit{Kepler} PSF relative to the aperture size, different apertures contain different portions of the flux from the target. This results in the severe discontinuities observed in the target flux. Fig. \ref{fig:kplr006932990-Lightcurves} shows the discontinuities present in the PDCSAP light curve of kplr006932990. The top panel is plotted in the observed frame where the sampling interval is 29.42 min. The bottom panel is plotted in the rest frame of kplr006932990 with sampling interval 28.64 min because of the cosmological redshift factor of $1+z = 1.0275$ in the case of this object. The effect of this cosmological time-contraction is to reduce the effective \textit{Kepler} sampling interval for more distant objects resulting in shorter but more densely sampled light curves. The discontinuities in the top panel of Fig. \ref{fig:kplr006932990-Lightcurves} must be removed before the long term variability properties of this object can be studied. One could remove such discontinuities by re-calibrating the data on an object-by-object basis, but this requires considerable manual effort. \textit{Kepler} data products include both the light curves of the target as well as the \textit{Target Pixel File} (TPF) corresponding to the target. The TPF consist of `postage stamp' snapshots of the object taken at every cadence. The light curve of the object is constructed by defining an aperture consisting of a subset of the pixels in the postage stamp. This aperture is determined by the \textit{Kepler} pipeline based on the \cite{bro11} and is optimized for stellar targets as outlined in \cite{DPH}. By re-defining the target aperture by hand, it is possible to reduce the discontinuities in the flux across quarters as shown in \cite{car12}. Such re-extracted light curves must then be de-trended of spacecraft induced features by removing the CBVs (discussed in \ref{KeplerMission}) with custom weights determined using the PyKE tool \textit{kepcotrend}. 

A simpler but cruder approach to removing quarterly discontinuities is to match a suitable metric across quarter boundaries \citep{kin12}. \cite{mus11} perform a simple end-matching to stitch light curves across quarters, i.e. they correct the measured flux values in the second of every sequential pair of quarters by the difference between the flux value of the last point of the leading member, and the flux value of the first point of the trailing member of the pair of quarters. This method is quite suitable for the high signal-to-noise (S/N) objects studied in \cite{mus11}, but does not work as well as the variability signal level drops closer to the instrument noise level in the dimmer targets. \cite{ede13} stitch the light curve of the BL Lac object kplr006690887 across quarters by using the per-quarter average flux value to determine a multiplicative factor that they then use to scale quarters. \cite{rev14} use a similar procedure but determine their multiplicative scaling factor from the average of the last 20 points in the quarter preceding the discontinuity and the first 20 points in the quarter following the discontinuity.

\begin{figure}
\center{\includegraphics[width=\columnwidth]{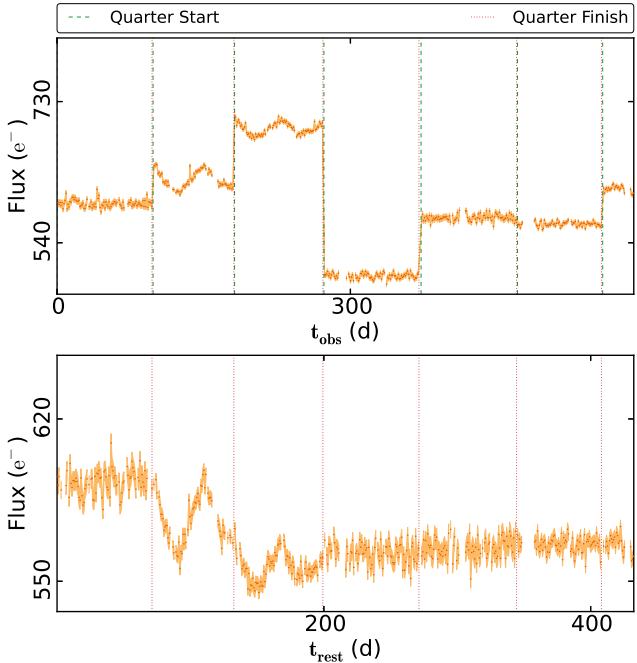}}
\caption{\label{fig:kplr003337670-Lightcurves} PDCSAP light curve of the Seyfert 1 kplr003337670 --- Top panel: unstitched; bottom panel: stitched. The noisiness of this light curve makes it essential to use a large (100) number of points at the edges of each quarters when performing the stitch. End-matching is completely ineffective while using a smaller number of points at the quarter edges results in obvious mis-matches across the quarter boundaries. }
\end{figure}

We stitch the light curves across quarterly discontinuities by matching the average flux value of the last 100 points of the leading quarter to the first 100 points of the trailing quarter in every sequential pair of quarters. The method is robust to outliers and is relatively unaffected by low S/N. It may be argued that it is more correct to use a fixed duration in time at the beginning and ends of quarters to compute average fluxes, rather than a fixed number of points. Using a variable number of points determined by a constant time window introduces variations in the reliability of the statistic for removing discontinuities; a one day window in the reference frame of a $z = 1$ object will have twice as many flux measurements than the same window in the case of a $z = 0$ object. Fixing the number of points has the benefit that it is equally applicable to stitching together the light curves of objects at unknown redshifts - the stitched light curves of such objects can stand on equal footing with the stitched light curves of objects with known redshift. Fig. \ref{fig:kplr006932990-Lightcurves} shows an application of our method to the high S/N light curve of kplr006932990. The top panel presents the un-stitched light curve while the bottom panel presents the stitched light curve corrected to the rest-frame of the AGN. As can be seen in the figure, our stitching method removes the discontinuities present in the original light curve. Similar results may be expected from the end-matching technique of \cite{mus11}. In contrast to the high S/N light curve of kplr006932990, Fig. \ref{fig:kplr003337670-Lightcurves} shows the un-stitched and stitched light curves of the low S/N AGN kplr003337670. The relative thickness of this light curve (caused by scatter from measurement noise) makes it essential to consider a suitably large number of points at the ends of quarters when stitching. Simple end-matching fails to stitch this light curve, and using a smaller number of points in the stitching results in obvious flux mis-matches. We use our stitching algorithm to remove the quarterly discontinuities in all 20 AGN light curves and discuss the resulting multi-quarter \textit{Kepler} light curves in the next section.

\section[]{AGN Light Curve Features}\label{StitchResults}

\begin{figure*}
\center{\includegraphics[width=\textwidth]{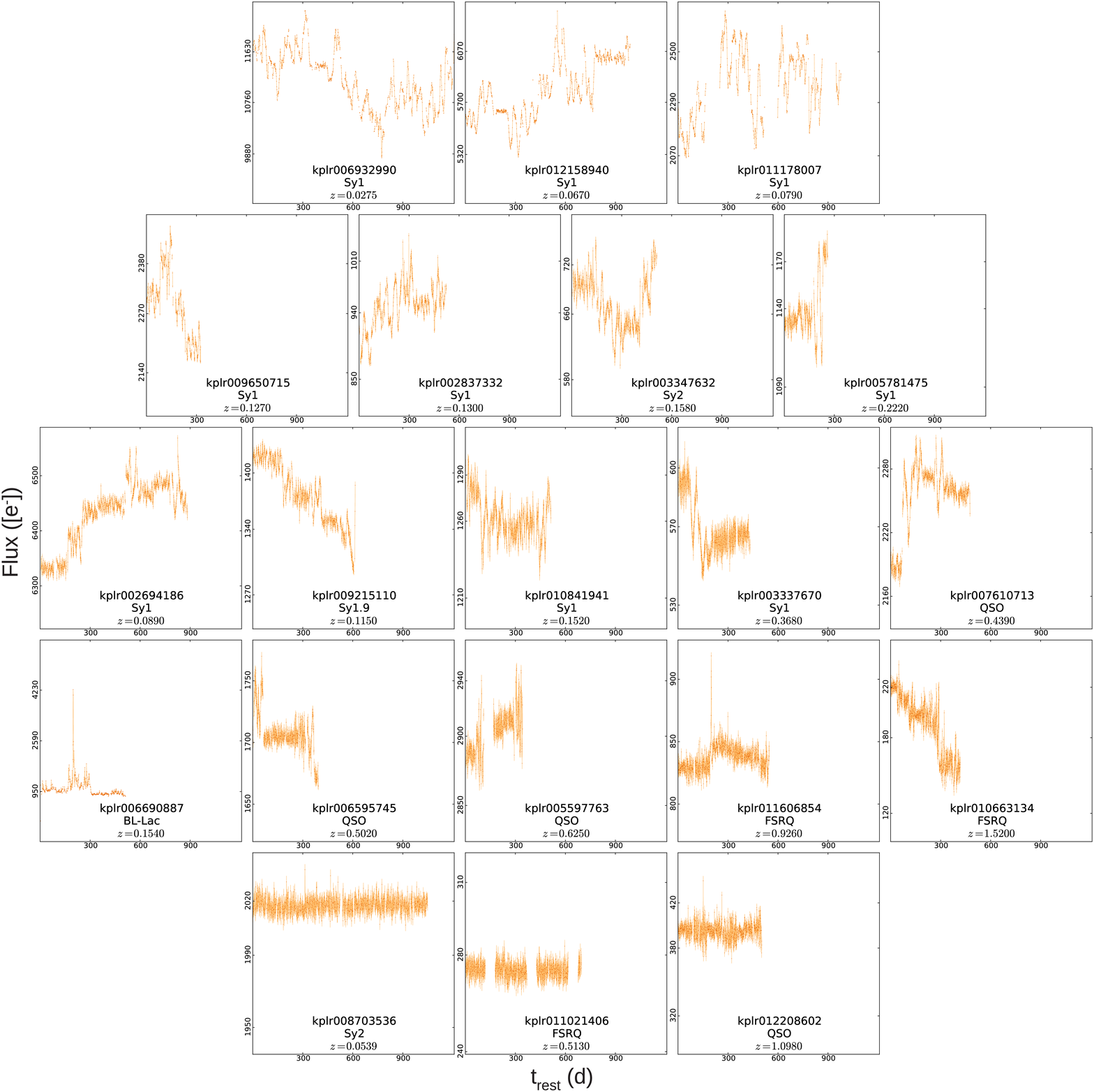}}
\caption{\label{fig:AllLightcurves} `Stitched' light curves of all the AGN in our investigation. The AGN are grouped based on the light curve categories discussed in Sec. \ref{StitchResults}. Light curves within a category are sorted in order of increasing redshift. All 20 light curves are plotted over the same rest-frame time range on the x-axis. Refer to Table \ref{tab:TargetList} for details on individual objects.}
\end{figure*}

We divide the AGN in the sample into 5 categories of objects based on the visual appearance of their light curve. Fig. \ref{fig:AllLightcurves} shows all 20 AGN grouped by visual features present in each light curve. Within each category, we order the AGN by $z$. All 20 light curves are plotted over the same rest-frame time range on the x-axis. Category 1 consists of 3 objects that have the most stochastic-looking light curves i.e. light curves bereft of any sinusoidal features or flares. Weak ripple features appear in the light curves of Category 2 objects (4 AGN). Category 3 consists of 5 objects that have pronounced ripple features in their light curves, while category 4 objects have light curves with flaring behavior (5 AGN). The last category of 3 objects have primarily featureless light curves consistent with marginal levels of variability.

The light curves show a wide variety of different types of behavior. There are indications that individual light curves can show more than one type of feature. Category 1 objects (kplr6932990 , kplr012158940, and kplr011178007) are shown in the 1$^{st}$ row of Fig. \ref{fig:AllLightcurves}. These objects have the most `stochastic'-looking light curves and are mostly free of any recurring features and trends. They are also the closest AGN in this study ($z < 0.1$), are all Seyfert 1s, and have the largest amplitudes in variability amongst all of the AGN in this study barring only kplr006690887.

Four AGN (kplr009650715, kplr002837332, kplr003347632, and kplr005781475) have mild oscillatory features in otherwise stochastic-looking lightcurves. This behavior is a mixture of the behaviors exhibited by the category 1 AGN above and the category 3 AGN discussed below. The light curves of these AGN are shown in the 2$^{nd}$ row of Fig. \ref{fig:AllLightcurves} between the light curves of the category 1 and 3 AGN to facilitate comparisons. Although there are indications of oscillatory behavior in this intermediate category of objects, the oscillations have very poorly defined time-periods as compared to objects in category 3. However, their light curves are not as purely stochastic looking as the light curves of the objects in category 1. Supporting this notion of an intermediate state is the observation that the object kplr012158940, which is grouped with the category 1 AGN, appears to be in the process of switching between categories as it begins to display features more reminiscent of category 3 light curves toward the end of the data. 

\begin{table}
\setlength{\tabcolsep}{4pt}
\caption{\label{tab:DetectorProps} \textit{Kepler} AGN Detector Sequences.}
\begin{tabular}{@{}rc@{}}
\hline
\hline
\multicolumn{1}{c}{Identifier} & \multicolumn{1}{c}{Detector Properties} \\
KeplerID & Module\#.Channel\#  \\
\hline
  6932990 & 14.48$^{\star}$, 8.24, 12.40$^{\ddagger}$, 18.64  \\
12158940 & 24.81$^{\star}$, 10.29, 2.1, 16.53 \\
11178007 & 23.79$^{\star}$, 15.51$^{\S}$, 3.7$^{\P}$, $11.35^{\S}$ \\
\hline
  9650715 & 14.46, 8.22$^{\dagger}$, 12.38, 18.62$^{\star}$\\
  2837332 & 22.74, 20.70, 4.10$^{\|}$, 6.14$^{\star}$  \\
  3347632 & 16.55, 24.83$^{\ddagger}$, 10.31, 2.3 \\
  5781475 & 7.17, 17.57, 19.65, 9.25  \\
\hline
  2694186 & 22.75$^{\star}$, 20.71, 4.11$^{\star}$, 6.15 \\
  9215110 & 13.41, 13.42, 13.43, 13.44$^{\ddagger}$$^{\dagger}$ \\
10841941 & 20.71, 4.11$^{\star}$, 6.15, 22.75$^{\star}$ \\
  3337670 & 16.53, 24.81$^{\star}$, 10.29, 2.1 \\
  7610713 & 8.22$^{\dagger}$, 12.38, 18.62$^{\star}$, 14.46 \\
\hline
  6690887 & 12.37, 18.61, 14.45, 8.21 \\
  6595745 & 8.22$^{\dagger}$, 12.38, 18.62$^{\star}$, 14.46 \\
  5597763 & 23.80, 15.52$^{\star}$, 3.8$^{\P}$, 11.36$^{\ddagger}$ \\
11606854 & 24.82, 10.30$^{\star}$, 2.2, 16.54 \\
10663134 & 19.66$^{\star}$, 9.26$^{\dagger}$, 7.18, 17.58$^{\ddagger}$$^{\dagger}$ \\
\hline
  8703536 & 9.27, 7.19, 17.59, 19.67 \\
11021406 & 23.77, 15.49, 3.5$^{\P}$, 11.33 \\
12208602 & 24.81$^{\star}$, 10.29, 2.1, 16.53 \\ 
\hline
\end{tabular}

$^{\star}$`medium' Rolling Band \& Moire

$^{\dagger}$`high' Rolling Band \& Moire 

$^{\ddagger}$Out of Spec Read Noise

$^{\S}$Out of Spec Undershoot

$^{\|}$Out of Spec Gain

$^{\P}$CCD Failed - no data

\medskip

The AGN are grouped based on light curve features as in Table \ref{tab:TargetList}. Col. 2 lists the module\#.channel\# sequence for each AGN. Note that some objects share this sequence of module\#.channel\# though the starting values may be different. This implies that such AGN are affected by similar levels of instrumentation artifacts and is used in Sec. \ref{StitchResults} to decide that instrumentation is not responsible for some of the features observed in the light curves.

\end{table}

\begin{figure}
\center{\includegraphics[width=\columnwidth]{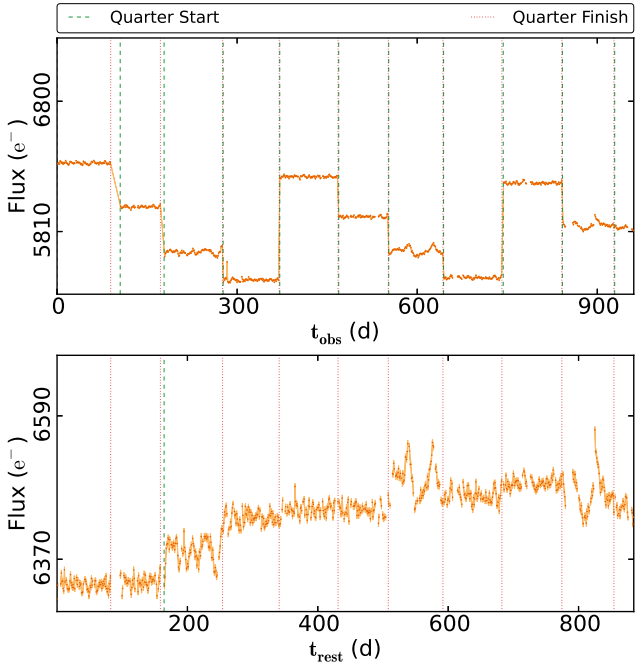}}
\caption{\label{fig:kplr002694186-Lightcurves} PDCSAP light curve of the Seyfert 1 kplr002694186 --- Top panel: unstitched; bottom panel: stitched. The oscillatory features in this light curve may at first be suspected to originate in the known \textit{Kepler} `moire' and `rolling-band' instrumentation effects. This AGN lands on the repeating module\#.channel\# pattern 22.75, 20.71, 4.11, and 6.15. While 22.75 and 4.11 are known to be prone to the moire and rolling-band problems, the other two are thought to be free of these effects suggesting that these oscillatory features are intrinsic to the light curve itself.}
\end{figure}

Category 3 AGN (kplr002694186, kplr009215110, kplr010841941, kplr003337670, and kplr007610713) are shown in the 3$^{rd}$ row of Fig. \ref{fig:AllLightcurves}. These objects appear to exhibit pronounced rippling features in the light curve. The aforementioned `moire' and `rolling-band' effects that are known to exist in specific detector modules and channels in \textit{Kepler} are the natural suspects. These effects are known to occur on timescales of a few days and have patterns very similar to those observed in these light curves \citep{kol10}. However, a close examination of the PDCSAP data suggests that the oscillations observed in these light curves are seen even when the target lands on a detector module and channel combination that is known to be free of the moire and rolling-band effects. For example, kplr002694186 exhibits strong oscillatory features in the light curve as seen in Fig. \ref{fig:kplr002694186-Lightcurves}. As reported in Table \ref{tab:DetectorProps}, this AGN falls on the module\#.channel\# combinations 22.75, 20.71, 4.11, and 6.15. Of these, 22.75 and 4.11 are known to be moderately affected by the rolling-band and moire artifacts, while 20.71 and 6.15 are thought to be free from these artifacts (\textit{Kepler} Instrumentation Handbook). Yet the same oscillatory signal is observed even during the quarters when the target falls on the un-affected module\#.channel\# combinations, strongly suggesting that the variations are intrinsic to the AGN. Further confirmation comes from the fact that kplr010841941 shares the same module\#.channel\# sequence as kplr002694186 but lags behind kplr002694186 by one quarter, suggesting that moire and rolling-band features seen in kplr002694186 should also be seen in kplr010841941 but lagging behind kplr002694186 by one quarter. No such phenomenon is observed, suggesting that the ripples are intrinsic to the AGN light curve rather than caused by instrumentation. In the case of kplr003337670, the module\#.channel\# sequence is 16.53, 24.81, 10.29, and 2.1. Of these only 24.81 suffers from moderate amounts of rolling-band and moire while the other 3 channels are clean. We observe (Fig. \ref{fig:kplr003337670-Lightcurves}) that the rapid low amplitude oscillations are present in the target during the first observed quarter when the target fell on the module\#.channel\# combination 16.53, which is thought to be clean. During the next quarter, on 24.81--a dirty module\#.channel\#--the AGN instead exhibits a lower-frequency, larger amplitude type oscillation. However, in the third observed quarter kplr003337670 falls on 10.29, a clean module\#.channel\#, while continuing to exhibit the same type of behavior. Such behavior cannot be ascribed to rolling-band or moire effects. Furthermore kplr003337670 shares module\#.channel\# sequence with kplr01215890 and kplr012208602, both of which show practically no rippling features in their light curve. The triplet of kplr007610713, kplr009650715, \& kplr006595745 all share the same sequence of module\#.channel\# but have very different looking light curves, suggesting that the moire and rolling-band effects are weak at best. Although these arguments do not conclusively prove that moire and rolling-band effects are not partly responsible for the ripple features observed in these light curves, we shall assume for the time being that these oscillatory features are real and postpone a more thorough investigation of residual data artifacts to the re-calibration of the light curves suggested in Sec \ref{Stitching}. 

The short-period low amplitude oscillations seen in category 3 AGN are punctuated by periods when the AGN appears to display a rather different sort of behavior that also has an oscillatory nature but is usually of larger amplitude and longer period, as described in the case of kplr003337670. Occasionally these punctuating periods result in overall changes in the flux level of the object. We caution that these oscillatory features may not be genuinely oscillatory, i.e. there are classes of ARMA random processes that can generate superficially oscillatory behavior given appropriate ARMA parameter choices--see examples in \cite{woo12}. 

\begin{figure}
\center{\includegraphics[width=\columnwidth]{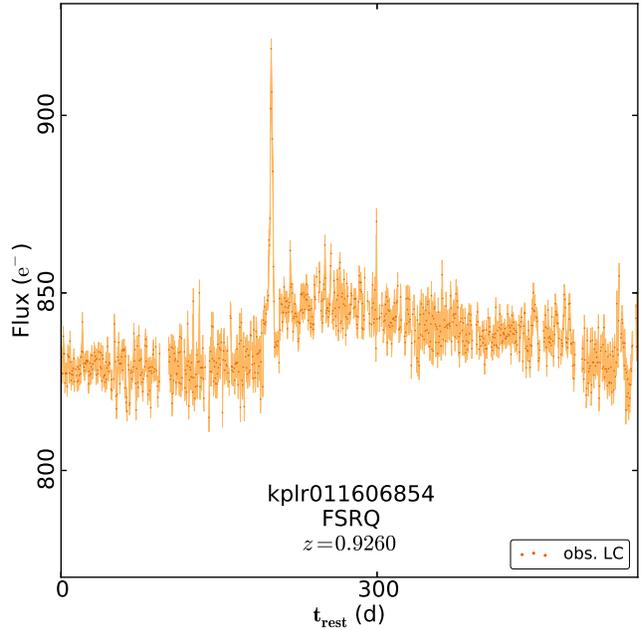}}
\caption{\label{fig:kplr011606854-Lightcurve} `Stitched' PDCSAP light curve of the FSRQ kplr011606854. After a featureless initial period, a burst of semi-oscillatory behavior occurs at the $\sim$ 75 d mark culminating in the flare at the $\sim$ 200 d mark. The light curve exhibits more oscillatory behavior after the flare but ultimately appears to settle back into an almost featureless phase by the $\sim$ 300 d mark suggesting that at least some AGN may display `on' and `off' phases in their light curve.}
\end{figure}

Category 4 AGN (kplr006690887, kplr006595745, kplr005597763, kplr0011606854, and kplr010663134) exhibit flares in their light curves as seen in row 4 of Fig. \ref{fig:AllLightcurves}. Three out of the five members of this class are blazars, implying that this type of behavior may be characteristic of blazars. However, no flares can be observed in the 4$^{th}$ blazar in our sample, kplr011021406 (which is also one of the objects that exhibits no variability). A solution to this puzzle may lie in the light curve of kplr011606854 in Fig. \ref{fig:kplr011606854-Lightcurve}. We see that after a featureless initial period, there is a burst of semi-oscillatory behavior at the $\sim$ 75 d mark culminating in the flare at the $\sim$ 200 d mark. The light curve exhibits more oscillatory behavior after the flare, but ultimately appears to settle back into an almost featureless phase by the $\sim$ 300 d mark, suggesting that at least some AGN may display `on' and `off' phases in their light curve.  

\begin{figure}
\center{\includegraphics[width=\columnwidth]{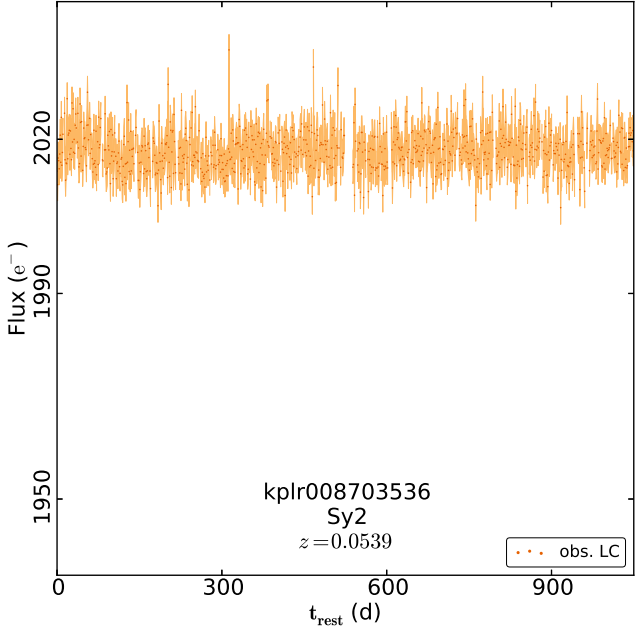}}
\caption{\label{fig:kplr008703536-Lightcurve} `Stitched' PDCSAP light curve of the Sy2 galaxy kplr008703536. We detect no variability signal in this object - the observed variations in this light curve may be ascribed to instrument noise suggesting that no endemic calibration problems exist in the dataset. }
\end{figure}

As mentioned earlier, not all of the objects show a measurable variability signal. Fig \ref{fig:kplr008703536-Lightcurve} shows stitched PDCSAP light curve of the Seyfert 2 galaxy kplr008703536. The variations seen in this light curve appear to be purely noise. Similar behavior is seen in the stitched light curve of the FSRQ/Sy 1.5 kplr011021406. The absence of features in these PDCSAP light curves and in a large number of stellar light curves available on the MAST suggests that the latest PDC module (Data Release 21 onwards) corrects most of the systematic errors in the \textit{Kepler} SAP light curves, leaving behind no endemic spacecraft event related artifacts. The only remaining sources of non-physical variability such as moire patterns, etc. must therefore be restricted to specific pixels and CCDs, allowing us to assume with high confidence that our stitched light curves are primarily astrophysical signal. The lack of variability observed in these two light curves confirms that not all AGN exhibit variability \citep{ses07}, though our sample is too small to be able to put constraints on the fraction of AGN that exhibit variability. The QSO kplr012208602 exhibits a very weak variability signal, occasionally displaying weak, intermittent semi-sinusoidal changes in flux. We posit that these changes would be un-observable using ground-based studies due to the weakness of the variability signal. The light curves for all 3 objects are plotted in the last row of Fig. \ref{fig:AllLightcurves}.  

We caution that per-channel/per-pixel effects may be responsible for some of the features observed in our light curves, particularly amongst the category 2 light curves. A true determination of the reality of some of these features will have to await the very systematic, customized per-object calibration discussed at the beginning of Sec. \ref{Stitching}. In the next section we discuss our method for analyzing the multi-quarter light curves.

\section[]{Structure Functions}\label{StructureFunctions}

We probe the variability properties of the \textit{Kepler} AGN light curves for consistency with the AR(1) process of Sec. \ref{AR1} by using structure functions to determine best-fit values of the parameter $\gamma$ in the damped power-law (DPL) model of Eq. \eqref{eq:DPL_PSD}. The DPL model is used to generate mock light curves with the same cadence and missing observation properties as the AGN under investigation. We estimate best fit model parameters by comparing the structure function of the observed light curve to the ensemble of structure functions computed from the mock light curves consistent with the model parameters. 

Many definitions of structure functions have appeared in astronomy literature (see \cite{gra14} for an overview). We use the definition provided by \cite{sim85} and \cite{ryt87}. The \textit{n-th Order Structure Function} of the process $F(t)$ is
\begin{equation}\label{eq:SF_DEF}
SF^{(n)}_{F}(\Delta t) = \langle \Xi^{(N)}_{F}(t,\Delta t)^{2} \rangle_{t},
\end{equation}
where $\Xi^{(N)}_{F}(t,\Delta t)$ is the \textit{N-th increment} of the process $F(t)$ at time-lag $\Delta t$. The N-th increment is given by
\begin{equation}\label{eq:Delta_DEF}
\Xi^{(N)}_{F}(t,\Delta t) = \sum_{m = 0}^{\infty} (-1)^{m} \left( \begin{array}{c} N \\ m \end{array} \right) F(t + [N-m]\Delta t).
\end{equation}
Using Eq. \eqref{eq:SF_DEF} and Eq. \eqref{eq:Delta_DEF}, the $1^{st}$ increment and $1^{st}$ Order Structure Function of the process $F(t)$ (hereafter referred to as `the increments' and `the SF') may be estimated using
\begin{equation}\label{eq:1Incr_DEF}
\Xi(t,\Delta t) = F(t + \Delta t) - F(t),
\end{equation}
and
\begin{equation}\label{eq:SF1_DEF}
SF(\Delta t) = \langle [F(t + \Delta t) - F(t)]^{2} \rangle.
\end{equation}
Structure functions offer benefits over directly estimating the PSD and ACVF/ACF of the process. Unlike the PSD, structure functions may be estimated in real space as opposed to Fourier space, making them more robust estimators of model parameters than PSD estimators that suffer from windowing and aliasing concerns \citep{rut78}. Structure functions offer an advantage over estimating the ACVF and ACF because of the de-trending properties of structure functions: an n-th Order Structure Function is insensitive to (n-1)-th order trends in the dataset. Unfortunately, the usefulness of this property is limited to the lower order structure functions. The \textit{Kepler} light curves are not long enough for computation of significantly higher order structure functions. Structure functions of all orders are related to the ACVF by simple linear equations. For example, the first order structure function can be related to the ACVF of $F(t)$ by
\begin{equation}\label{eq:SF-ACF_Relation}
SF(\Delta t) = 2ACVF(0) - 2ACVF(\Delta t).
\end{equation}

\begin{figure}
\center{\includegraphics[width=\columnwidth]{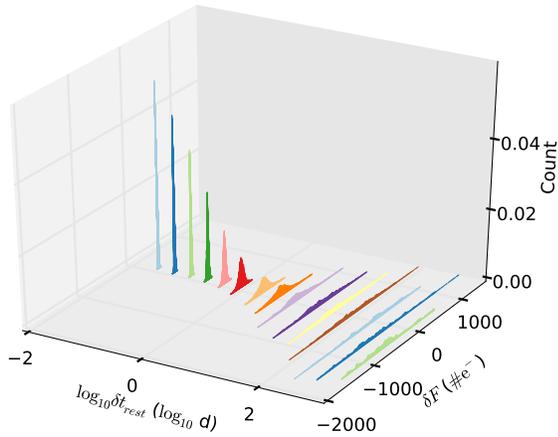}}
\caption{\label{fig:kplr006932990-1stIncr3D} Histogram plots of the $1^{st}$ increment ($F(t + \Delta t) - F(t)$) for various choices of the time-lag $\Delta t$ drawn from the light curve of the Seyfert 1 galaxy kplr006932990. Each histogram shows the distribution of the $1^{st}$ increment values for a given choice of time lag $\Delta t$. The $1^{st}$-order structure function is the variance of the distribution in each histogram and quantifies how the variance of the increments changes as a function of time lag.}
\end{figure}

To understand what information the structure function conveys, consider the histograms of flux differences, i.e. $1^{st}$ increments for various values of time-lag $\Delta t$ shown in Fig. \ref{fig:kplr006932990-1stIncr3D}. By definition, the structure function at time-lag $\Delta t$ is the average of the squares of the values entering each histogram, i.e. it is the variance of the flux differences at the chosen time-lag $\Delta t$. The structure function therefore quantifies how the variance of the flux differences changes as a function of time-lag. In the case of objects that show stochastic variations in $F(t)$, the histograms look like bell-shaped curves at small time-lags. As the time-lag increases, the width of the bell-curves increase until a critical time-lag $\tau$ is reached. In the case of the AR(1) process, this critical time-lag can be identified with the the turnover timescale in Eq. \eqref{eq:AR1}.

Astronomical time series have measurement noise that is usually modeled as uncorrelated white-noise, i.e. $ACVF(0) = \sigma^{2} + \sigma_{N}^{2}$ where $\sigma^{2}$ is the contribution to the variance of the time series at zero-lag from the actual signal, while $\sigma_{N}^{2}$ characterizes the noise level of the measurement noise. In the case of the AR(1) process, the ACVF takes the form in Eq. \eqref{eq:DRW_ACF}. If the variability observed in \textit{Kepler} AGN light curves is well described by an AR(1) process, then the structure function should be well described by
\begin{equation}\label{eq:DRW_SF}
SF_{DRW}(\Delta t) = 2 \sigma^{2} (1 - e^{- \left| \frac{\Delta t}{\tau} \right|} ) + 2 \sigma_{N}^{2}.
\end{equation}
This may be tested by estimating the structure function of a real AGN light curve to check if it is consistent with Eq. \eqref{eq:DRW_SF} or with a more general form based on the DPL PSD of Eq. \eqref{eq:DPL_PSD}. Unfortunately there is no closed-form expression for the auto-correlation function corresponding to the DPL PSD, making it impossible to directly fit the functional form for the structure function observed for a given AGN. In the next section, we describe a Monte-Carlo estimation technique for recovering the DPL model parameters.

\section[]{Monte-Carlo Estimation of the Damped Power-Law Model Parameters}\label{Estimation}

\begin{figure}
\center{\includegraphics[width=\columnwidth]{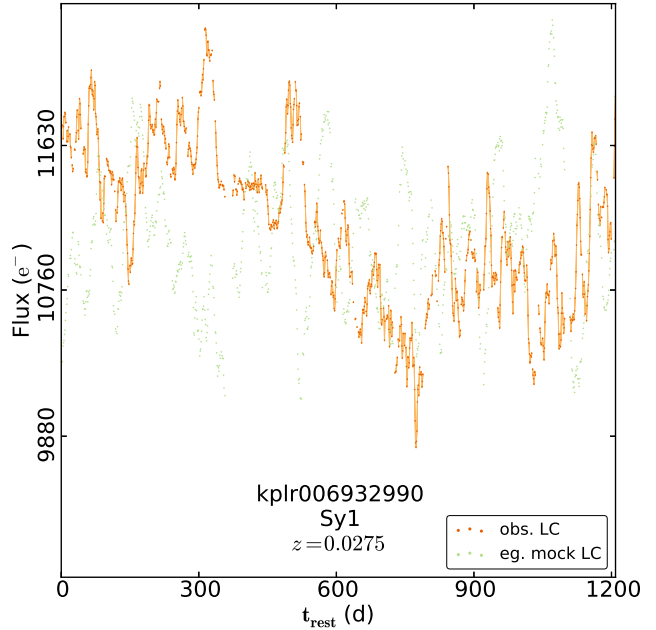}}
\caption{\label{fig:kplr006932990-mockLC} Observed light curve of the Sy 1 AGN kplr006932990 (orange) along with an example mock light curve (light green) generated as described in \ref{Estimation} (generated with $\gamma = 2.7$ and $\tau = 27.5$ d). The mock light curves used in Monte-Carlo estimation process have the exact sampling pattern of the observed light curve ensuring that the structure functions of the mocks are subject to the same sampling issues as that of the real galaxy.}
\end{figure}

We estimate the DPL parameters in Eq. \eqref{eq:DPL_PSD} via Monte-Carlo simulations. We compare the real structure function of an AGN to simulations computed from mock light curves generated using the DPL model of Eq. \eqref{eq:DPL_PSD}. We generate `mock' light curves using the \cite{tim95} method. To create a single mock light curve, pseudo-random numbers are generated using the Fast Mersenne Twister SFMT19937 generator seeded with hardware-generated random numbers (generated using Intel RDRAND instruction) to ensure that the random number sequences are free of artificial correlations \citep{cod94} induced by poor random seed choices. At this intermediate stage, the mock light curve is oversampled by a factor of 10 i.e. we generate points at $10 \times$ the required cadence in order to avoid sampling issues. To include low frequency modes that are not adequately characterized by the length of the observed light curve, the intermediate mock light curve is much longer than is ultimately required to make the final mock; mock light curves generated in this manner are capable of exhibiting low frequency modes longer than the length of the observed data. FFTs are most efficient for data sequences that are a power of 2; for this reason, we pick the intermediate (including the oversampling) to be of length $2^{23}$. This results in the intermediate mock light curve being between $15 \times$ to $45 \times$ the length required for the final mock light curve depending on the actual length of the observed light curve. We pick a uniformly distributed random segment of the intermediate overly-long light curve that has the same length as the observed light curve and generate another stream of un-correlated Gaussian random deviates to simulate the white-noise properties of \textit{Kepler} instrumentation noise. After adding this `measurement noise', we set data points corresponding to the un-observed cadences in the real light curve to $0$. This procedure creates a final mock light curve with identical sampling and noise properties to the real light curve. Fig. \ref{fig:kplr006932990-mockLC} shows the true light curve (orange) along with an example mock light curve (light green) for the Sy 1 AGN kplr006932990 illustrating what the mock light curves look like for the best fit DPL parameters for this object. 

We compute the structure functions using Eq. \eqref{eq:SF1_DEF} modified to account for missing values -
\begin{equation}\label{eq:SF1_DEF_MOD}
SF_{n} = \frac{\sum_{i = 1}^{N - n}w_{i}w_{i+n}[F_{i + n} - F_{i}]^{2}}{\sum_{i = 1}^{N - n}w_{i}w_{i+n}},
\end{equation}
where $n = \Delta t/\delta t$ is the lag in cadences as opposed to physical time. The weights are defined to be $w_{i} = 1$ for observed cadences and $w_{i} = 0$ otherwise.

\begin{figure}
\center{\includegraphics[width=\columnwidth]{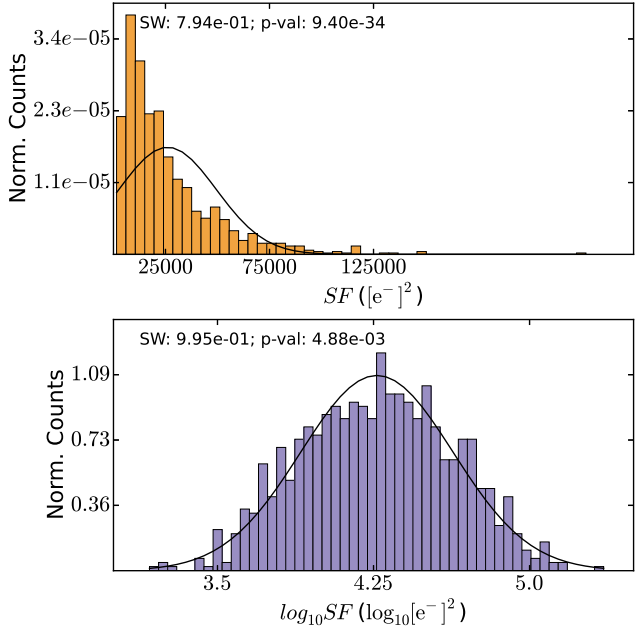}}
\caption{\label{fig:SFDistribution} Distribution of mock $SF(\delta t)$ estimates for $1000$ simulated light curves of length $T = 10$ d, with $\gamma = 2.0$ \& $\tau = 7.5$ d for $\delta t = 5$ d. The upper panel shows the (clearly non-Gaussian) distribution of $SF(\delta t)$, while the lower panel shows the distribution of $\log_{10} SF(\delta t)$. The Shapiro-Wilk test W- \& p-values confirms that the distribution of $\log_{10} SF(\delta t)$ is closer to Gaussian than that of the raw $SF(\delta t)$.}
\end{figure}

It is known that, for any given lag $\Delta t$, the distribution of $SF(\Delta t)$ generated using the Timmer-K\"{o}nig method is not Gaussian \citep{emm10}. However, the logarithms of the structure function, $\log_{10}SF(\Delta t)$, are distributed as per a Gaussian distribution. As can be seen in Fig. \ref{fig:SFDistribution}, histograms of mock $SF(\Delta t)$ and $\log_{10}SF(\Delta t)$ values for a set of 10000 mock light curves constructed with PSD slope $\gamma = 2$, and characteristic timescale $\tau = 7.5 d$ at $\Delta t = 5 d$ for mock light curves of length $T = 10$ d suggest that while the estimates of $SF(\Delta t)$ are not Gaussian distributed, the estimates of $\log_{10} SF(\Delta t)$ are. This observation may be confirmed using a test such as the Shapiro-Wilk test. Note however, that even though the Shapiro-Wilk test confirms that $\log_{10} SF(\Delta t)$ is closer to Gaussian than $SF(\Delta t)$, it is not always Gaussian in an absolute sense; a more through investigation of the distribution properties of the structure function may be warranted. We use $\log_{10} SF_{n}$ instead of $SF_{n}$ to compare structure functions between observations and simulations.

Since the logarithms of the structure function estimates are drawn from Gaussian distributions, we can use the mean and standard deviation vectors of the logarithms of the mock structure functions to estimate the best-fit DPL model parameters. We compute the mean $\mu_{n} = \langle \log_{10}SF_{n} \rangle$ and variance $\sigma^{2}_{n} = \langle (\log_{10} SF_{n} - \langle \log_{10} SF_{n} \rangle )^{2} \rangle$ of the ensemble of the logarithms of the mock structure functions and numerically minimize
\begin{equation}\label{eq:Chi2}
\chi^{2} = \sum_{n = 1}^{N-1} \frac{(\mu_{n} - \log_{10} SF_{n,obs})^{2}}{\sigma_{n}^{2}},
\end{equation}
using the Constrained Optimization BY Linear Approximations (COBYLA) optimization algorithm \citep{pow94} to search the parameter space for the best-fit estimates of $\gamma$, $\tau$, $\sigma_{S}$, and $\sigma_{N}$. Coarse optimization is performed using $10^{4}$ mocks at each step in the optimization, after which the best-fit parameter estimates are refined using $10^{5}$ mocks at each step in the optimization process. Such large numbers of mocks are required to ensure that $\chi^{2}$ values computed at the same point in parameter space (using different choices of random seeds) are within the tolerances chosen in the optimization step.

The resulting values of the reduced chi-square per degree-of-freedom, $\chi^{2}_{DoF}$ are close to $1$. We also compute the percentile, $\mathsf{P}$, of the value of $\chi^{2}_{DoF}$ corresponding to the best-fit parameter values using a fresh set of $1000$ mock light curves. Percentile values $\mathsf{P}$ close to $100 \%$ mean that the values of $\chi^{2}_{mock}$ are almost uniformly smaller than $\chi^{2}_{obs}$ for the best-fit model parameters, indicating that the model is a poor fit to the data.   

\begin{figure}
\center{\includegraphics[width=\columnwidth]{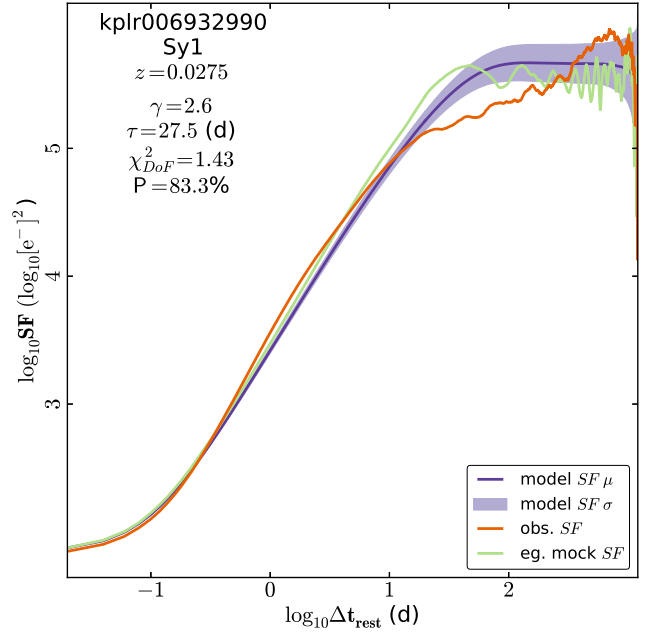}}
\caption{\label{fig:kplr006932990-SF1} Observed structure function of the Sy 1 AGN kplr006932990 (orange), best-fit model structure function with standard deviations (purple) and the mock structure function (light green) corresponding to the mock light curve in Fig. \ref{fig:kplr006932990-mockLC}. The value of $\gamma$ is most sensitive to the short timescale ($\Delta t \lesssim 5$ d) behavior of the structure function where the variance allowed is small.}
\end{figure}

Fig. \ref{fig:kplr006932990-SF1} shows the observed structure function of the Seyfert 1 kplr006932990 (orange) along with the best-fit mock structure function and standard deviation (purple) and the mock structure function (light green) corresponding to the mock light curve in Fig. \ref{fig:kplr006932990-mockLC}. On short timescales ($\Delta t  \lesssim 5$ d), the mock structure functions show very small standard deviation, while on longer timescales, the structure function is allowed to vary more significantly. The value of $\gamma$ is determined almost exclusively by this short timescale ($\Delta t \lesssim 5$ d) behavior of the structure function where the variations allowed in the possible structure function realizations are small. Note that in the case of light curves that have $\gamma > 2.0$, as the PSD turns over from shorter timescales ($\Delta t \lessapprox 5$ d with local PSD slope $\gamma > 2$) to longer timescales ($\Delta t \gtrapprox 10$ d with local PSD slope $\gamma \rightarrow 0$), the \textit{local} value of the PSD slope passes through $\gamma \equiv 2$. Most ground-based surveys, such as the SDSS, have large (compared to \textit{Kepler}) photometric errors and are measuring the structure function on timescales $\Delta t \gtrapprox 5$ d resulting in estimates of $\gamma \sim 2$. 

Ideally one would estimate confidence intervals for the model parameters using algorithms such as MCMC to sample the parameter. Unfortunately, the Monte Carlo simulation is computationally very expensive with the calculation of $\chi^{2}$ taking $\sim 3$ h for every location in parameter space on a $16$-core Xeon machine with $2$ Xeon Phi accelerator cards. This makes the Monte-Carlo technique of fitting the structure function unsuitable for an MCMC style analysis. To estimate the error in our computation of the PSD slope $\gamma$ we generated simulations of short segments of mock light curves with known DPL model parameters and then proceeded to recover the input parameters. Based on this, we estimate that the error in the value of $\gamma$ obtained is on the order of $10 \%$.

\section[]{Determination of $T_{min}$}\label{TMin}

\begin{figure}
\center{\includegraphics[width=\columnwidth]{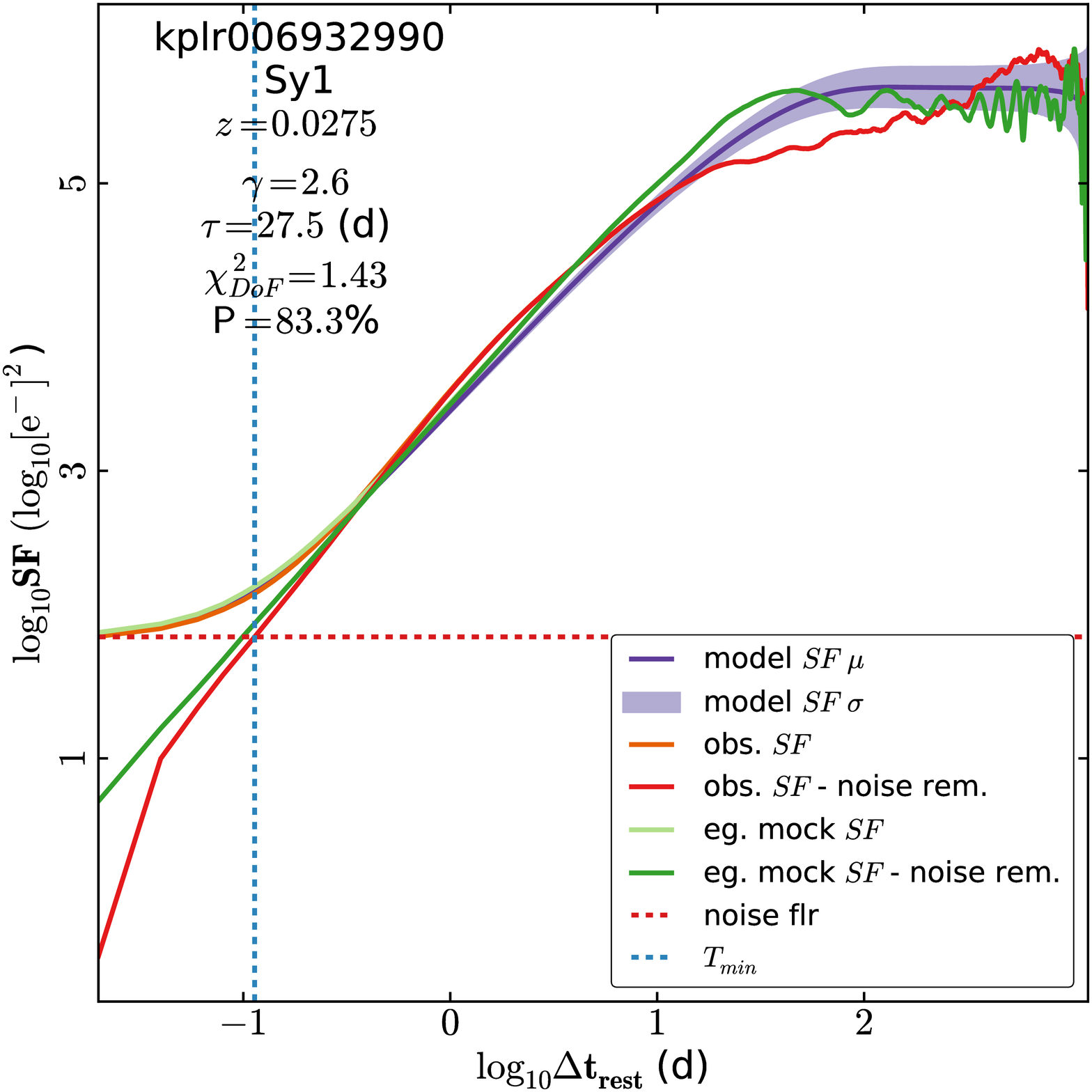}}
\caption{\label{fig:kplr006932990-SF1+Timescale} Observed structure function of the Sy 1 AGN kplr006932990 (orange) and the best-fit model structure function with standard deviation (purple). The noiseless structure function (red) is obtained by removing the noise floor (red dashed-line) after which the shortest variability timescale $T_{min}$ (blue) is determined from the intersection of the noiseless structure function with the noise floor. $T_{min}$ typically $< 1 $ d implies that future AGN variability studies should attempt to sample the short timescale variability properties of AGN.}
\end{figure}

We estimate the shortest timescales on which we observe variability in the light curves. Our estimation of the DPL model parameters in Sec. \ref{Estimation} yields an estimate of the un-correlated white-noise level present in each light curve. This noise level can be seen in the structure function in the form of a flattening-out of the structure function on very short timescales. For example, in Fig. \ref{fig:kplr006932990-SF1+Timescale}, we observe that the structure function of the Seyfert 1 kplr006932990 begins to level off to a value of $\sim 100 [\mathrm{e^{-}}]^{2}$ on timescales of $\sim 0.5$ d. From Eq. \eqref{eq:DRW_SF}, we expect that this is caused by the noise level i.e. $2\sigma_{N}^{2} \sim 100 [\mathrm{e^{-}}]^{2}$. By removing this noise level from the observed structure function, we can estimate a noiseless version of the structure function. To estimate the shortest timescale on which we observe variability in the light curve, we find the time-lag at which this noiseless structure function crosses the noise floor. In Fig. \ref{fig:kplr006932990-SF1+Timescale} we plot the observed noisy structure function of kplr006932990 (orange), our best fit estimates of this structure function (purple), and the noise level determined from the structure function fit (red dashed-line). After removing this noise level from the observed structure function, we obtain the noiseless structure function (red). We also show the mock structure function (light green) corresponding to the mock light curve in Fig. \ref{fig:kplr006932990-mockLC} along with the noiseless version (green) of this mock structure function. We define the variability onset timescale $T_{min}$ to be the intersection point (indicated by blue dashed-line) of the noiseless structure function with the noise floor. We compute the value of the variability onset timescale for the AGN in this study along with the rest-frame sampling interval for each object and present the results in Table \ref{tab:Results}.  

\section[]{Observed Structure Functions and Fits}\label{Results}

\begin{figure*}
\center{\includegraphics[width=\textwidth]{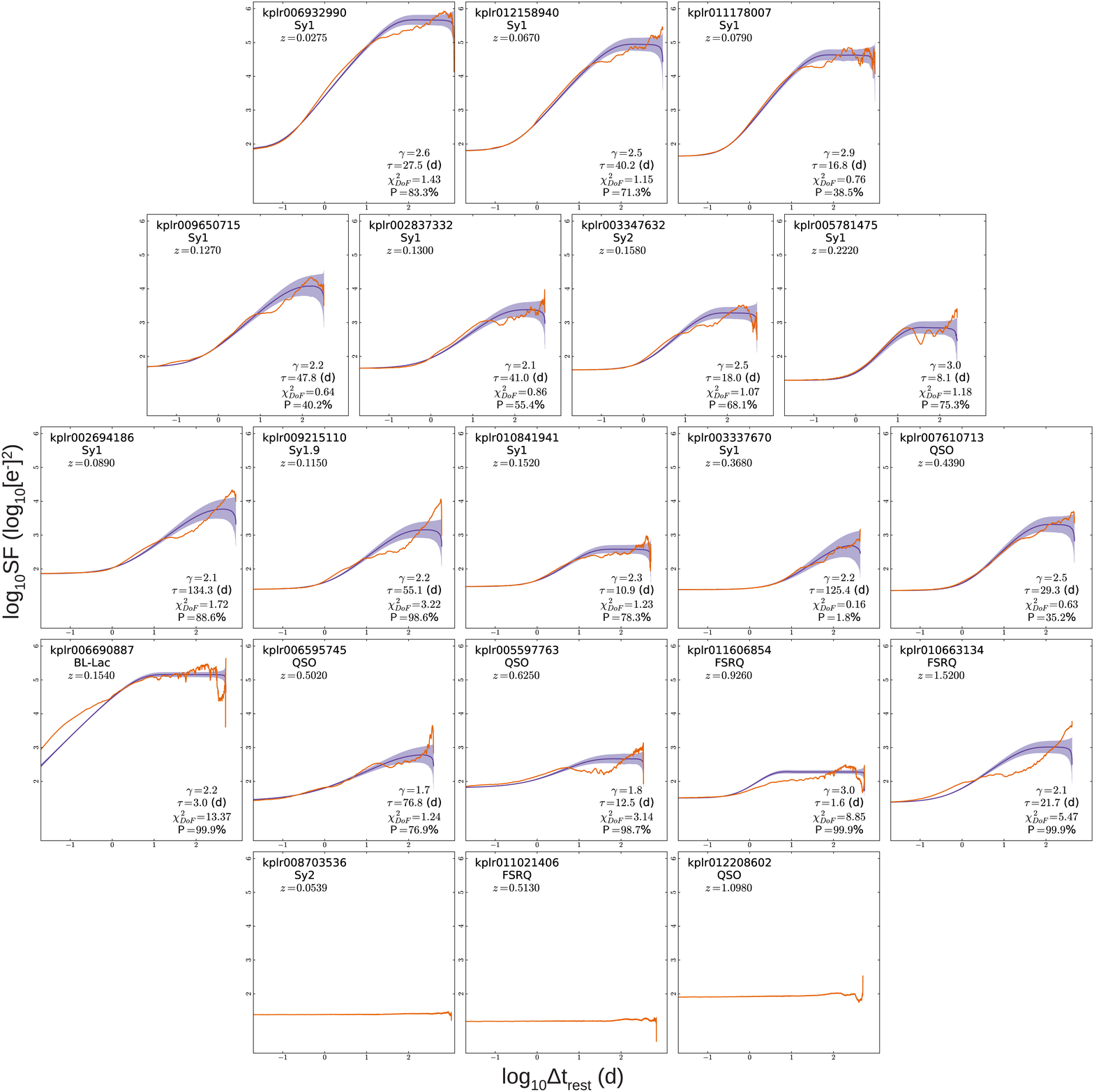}}
\caption{\label{fig:StructureFunctions} Observed structure functions (orange) and fits using mock light curves (purple) of all the AGN. Note that all the structure functions have been plotted at the same scale. }
\end{figure*}

The excellent short timescale sampling properties of \textit{Kepler} make it possible to study the AGN structure functions with unprecedented levels of detail. Using the Monte-Carlo fitting method of Sec. \ref{Estimation}, we fit the DPL parameters of Eq. \eqref{eq:DPL_PSD} for the 17 most variable objects. Fig. \ref{fig:StructureFunctions} shows the observed structure functions (orange), Monte-Carlo fits (purple line), and 1$\sigma$ variation (purple shaded region).  

Sampling information and estimates of the DPL parameters are given in Table \ref{tab:Results} for each \textit{Kepler} AGN. Column 1 lists the \textit{KeplerID} of the object. Column 2 lists the length of each light curve in quarters. Each quarter spans about $3$ months in duration and has $\sim 4300$ data points. Included in this figure are quarters during which the target fell on a failed CCD module, resulting in one or more missing quarters of data if flux measurements exist before and after the `missing' quarter. Column 3 lists the rest-frame sampling interval for each object in minutes. The next two columns list the best-fit values for the DPL parameters $\gamma$ and $\tau$ from Eq. \eqref{eq:DPL_PSD} where $\tau$ is measured in rest-frame days. Column 6 lists the best-fit estimate for the minimum timescale on which variability is observed $T_{min}$ in minutes. The `goodness-of-fit', i.e. the reduced chi-square, is listed in Column 7 while Column 8 lists the percentile of the chi-square of the observed structure function fit as compared to simulations drawn from the DPL model i.e. it is also an indicator of `goodness-of-fit' in that a very large value of the chi-square percentile indicates that the DPL model is insufficient to explain the data.

\begin{table}
\setlength{\tabcolsep}{4pt}
\caption{\label{tab:Results} Sampling Parameters \& Results.}
\begin{tabular}{@{}rcrccccc@{}}
\hline
\hline
\multicolumn{1}{c}{Identifier} & \multicolumn{2}{c}{Sampl. Param.} & \multicolumn{5}{c}{Analysis Results} \\
KeplerID & L & $\delta t_{rest}$ & $\gamma$ & $\tau$ & $T_{min}$ & $\chi^{2}_{DoF}$ & $\mathsf{P}$ \\
 & (\#Q) & (min) & & (d) & (min) & & \% \\
\hline
  6932990 & 14 & 28.64 & 2.7 &   27.5 &   163 & 1.43 & 83.3 \\
12158940 & 12 & 27.58 & 2.5 &   40.2 &   440 & 1.15 & 71.3 \\
11178007  & 11 & 27.27 & 2.9 &   16.8 &   428 & 0.75 & 38.5 \\
\hline
  9650715 &    4 & 26.11 & 2.2 &  47.8 &   502 & 0.64 & 40.2 \\
  2837332 &    7 & 26.04 & 2.1 &  41.0 & 1358 & 0.86 & 55.4 \\
  3347632 &    7 & 25.41 & 2.5 &  18.0 & 1453 & 1.07 & 68.1 \\
  5781475 &    4 & 24.08 & 3.1 &  81.0 & 1619 & 1.18 & 75.3 \\
\hline
  2694186$^{\star}$ &  11 & 27.02 & 2.1 &  134.3 & 2339 & 1.72 & 88.6 \\
  9215110 &    8 & 26.39 & 2.2 &   55.1 & 1704 & 3.22 & 98.6 \\
10841941 &    7 & 25.54 & 2.3 &   10.9 & 2151 & 1.23 & 78.3 \\
  3337670 &    7 & 21.51 & 2.2 & 125.4 & 9946 & 0.16 & 1.80 \\
  7610713 &    8 & 20.45 & 2.5 &   29.3 & 1677 & 0.63 & 35.2 \\
\hline
  6690887 &    7 & 25.50 & 2.2 &     3.1 &             - & 13.37 & 99.9 \\
  6595745 &    7 & 19.59 & 1.7 &   76.8 &   711 &   1.23 & 76.9 \\
  5597763 &    7 & 18.11 & 1.8 &   12.5 & 1151 &   3.14 & 98.7 \\
11606854 &  12 & 15.28 & 3.0 &     1.6 & 2005 &   8.85 & 99.9 \\
10663134 &  12 & 11.68 & 2.1 &   21.7 &   463 &     5.47 & 99.9 \\
\hline
  8703536 & 12 & 27.92 &        - &            - &              - & - & - \\
11021406 & 12 & 19.45 &        - &            - &              -  & - & - \\
12208602 & 12 & 14.02 &        - &            - &              - & - & - \\ 
\hline
\end{tabular}

$^{\star}$kplr002694186 has a total light curve spanning 17 quarters in the MAST database. Data is available only for quarter 0 (as an exo-planet search program target) and then subsequently from quarters 7 through 17 (as a GO program target). We discard the quarter 0 light curve segment because of the unreliability of the photometry during this initial phase of the mission and report kplr2694186 as having a light curve of length 11 quarters. 

\medskip

The AGN are grouped based on light curve features as in Table \ref{tab:TargetList}. Column 2 (`\# Q') is the approximate length of each light curve in quarters. Note that occasionally a target will land on one of the failed CCD modules and will not have data available for the quarters corresponding to such occurrences. Column 2 includes such `missing' quarters since the overall length of the light curve just depends on the first and last quarter observed. For example, the radio-loud AGN  kplr011021406 does not have data for quarters 8, 12, and 16 but since this object was observed starting in quarter 6 and ending in quarter 17, the light curve still has a total length of 12 quarters of data. Other than the incomplete 17th quarter, each quarter contains approximately 4300 individual data points making the longest light curve (kplr006932990) over 60,000 points long. Column 3 lists the rest-frame sampling interval for each object. Columns 4 through 6 list the results of the structure function analysis where $\gamma$ is the best fit slope of the power-law portion of the power spectral density, $\tau$ is the turnover timescale beyond which individual data points are essentially un-correlated, and $T_{min}$ is the shortest timescale on which variability is observed. Columns 7 \& 8 list measures of the `goodness-of-fit' i.e. the reduced $\chi^{2}$ and the percentile value $\mathsf{P}$ of the chi-square compared to chi-squares computed from 1000 mock light curves.
\end{table}

We observe that in all but one case (the blazar kplr00669887), \textit{Kepler} samples the light curve on shorter sampling intervals $\delta t_{rest}$ (Col. 4 in Tab. \ref{tab:Results}) than $T_{min}$ (Col. 7). The median variability onset timescale is $\sim 23$ h in the rest frame of the object, while the minimum and maximum values are $\sim 2.5$ h and $\sim 1.6$ d respectively. The structure function of kplr006690887 is notable for being bereft of a well-defined noise floor implying that this BL Lac object varies even on the 25 min timescale probed by \textit{Kepler} in the rest-frame of this object. This suggests that the \textit{Kepler} sampling pattern is more than adequate to study optical AGN variability. We caution that $T_{min}$ is merely an upper estimate for the shortest timescale on which variability can be observed given the noise level, therefore even better photometric precision than is available with \textit{Kepler} is required to find the true shortest timescale on which AGN vary.

We categorize the light curves of our AGN based on the presence of oscillatory features and flares in Sec. \ref{StitchResults}. All 3 objects that we grouped in category 1 (completely stochastic-looking light curves--row 1 of Fig. \ref{fig:AllLightcurves}) are fit by DPL models with median $\gamma = 2.7$ with $\chi^{2}_{DoF}$ close to $1$. This value of $\gamma$ is significantly different from that expected of an AR(1) process and indicates that the light curves of these 3 objects are smoother than a damped random walk ($\gamma_{DRW} = 2.0$). Prominent in the observed structure function fits for these three AGN in the 1$^{st}$ row of Fig. \ref{fig:StructureFunctions} is the presence of a `dip' feature on time-lags $\Delta t$ between $\sim 10$ and $100$ d. This dip feature may be interpreted as an excess of correlation on these timescales. The dip starts at $\sim 10$ d in the \textit{rest-frame} of the objects, suggesting that it is probably intrinsic to the light curves of these objects. Such dips rarely occur in the structure functions of the simulated light curves (generally these exhibit wiggling behavior after the turnover timescale is reached). This suggests that, even though all three objects have $\chi^{2}$ values consistent with the DPL model ($\mathsf{P}$ ranges between $40$\% and $80$\%), the DPL model is probably too simple to characterize the variability observed in these objects.

Of the objects categorized as intermediate between categories 1 and 3 (stochastic-looking light curves+oscillatory features--row 2 of Fig. \ref{fig:AllLightcurves}), the first 2 have PSD slopes ($\gamma = 2.2$ and $\gamma = 2.1$ respectively) close to that expected from an AR(1) process. Of these first two objects, kplr009650715 has a fairly low $\chi^{2}_{DoF} = 0.64$ implying that the DPL model may be overfitting the light curve of this object. The latter two objects, kplr003347632 \& kplr5781475, have much steeper PSD slopes ($\gamma = 2.5$ and $\gamma = 3.1$ respectively) completely inconsistent with the AR(1) process and are well fit by the DPL model. The median PSD slope for this category is $\gamma = 2.35$ and the individual structure functions are shown in row 2 of Fig. \ref{fig:StructureFunctions}.  

\begin{figure}
\center{\includegraphics[width=\columnwidth]{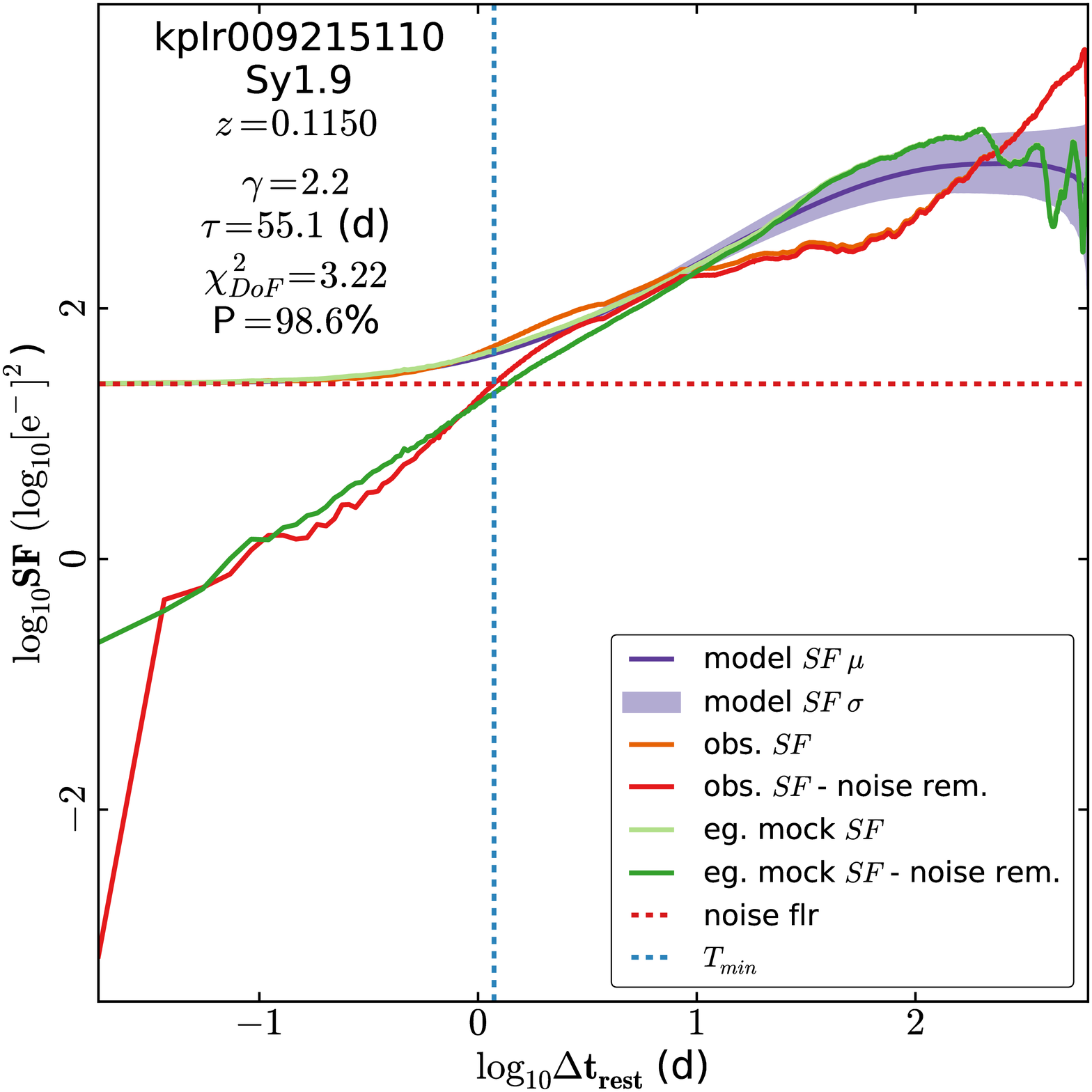}}
\caption{\label{fig:kplr009215110-SF1} Observed structure function of the Sy 1.9 AGN kplr009215110 (orange), noise-removed observed structure function (red), example mock structure function (light green), noiseless mock structure function (green), and best-fit model structure function with standard deviations (purple). This structure function exhibits varying PSD slope $\gamma$. The structure function rises steeply when $\Delta t_{rest} \lesssim 3$ d after which is flattens somewhat. The DPL model is incapable of reproducing such behavior.}
\end{figure}

Objects that fall into category 3 (strong oscillatory features in light curves--row 3 of Fig. \ref{fig:AllLightcurves}) based on a visual inspection of the light curve have the largest number of light curve slopes consistent with that expected from the AR(1) process. Four out of 5 have $\gamma \sim 2.2$, while the median estimate of $\gamma = 2.2$ for all 5 combined. Individual estimates of $\gamma$ range between $2.1$ (kplr002694186) and $2.5$ (kplr007610713). Even though our structure function estimation algorithm produces estimates of the turnover timescale $\tau$ for all of these objects, a visual inspection of the corresponding structure functions in the row 3 of Fig. \ref{fig:StructureFunctions} suggests that the turnover timescale is suspect in all but the case of kplr010841941, as borne out by the mock structure function error contours. Dips similar to those observed in category 1 objects are also seen in the structure function of some of these objects. As in the case of category 1 AGN, the dip feature begins around $\sim 10$ d in the \textit{rest-frame} of the objects and is usually absent by $\sim 100$ d. kplr002694186 \& kplr009215110 exhibit structure functions that appear to have different slopes on different timescales. This presence of multiple slopes may be responsible for the relatively poor quality of the fits of the structure functions of these objects ($\chi^{2}_{DoF}$ is $1.72$ and $3.22$ respectively). Fig. \ref{fig:kplr009215110-SF1} shows the structure function of the Sy 1.9 AGN kplr009215110. On short timescales $\lesssim 3$ d the structure function rises steeply, after which it flattens somewhat. The DPL model is incapable of producing such behavior. Similar behavior is also observed in kplr006595745.

Category 4 objects with flares in the light curves exhibit a broad spread in estimated $\gamma$ ranging from a fairly flat $1.7$ (kplr006595745) to a very steep $3.0$ (kplr011606854). However, this category of objects have the worst fits as a class, with very suspicious $\chi^{2}_{DoF}$ values such as $13.4$, $8.9$, \& $5.5$ (kplr006690887, kplr011606854, \& kplr010663134). All of these objects have much higher $\chi^{2}$ values than is the norm for mock light curves generated using the DPL model, suggesting that the DPL model (and therefore the AR(1) process) are wholly unsuitable for modeling the light curves of objects that show flaring behavior. 
      
The last row of Fig. \ref{fig:StructureFunctions} shows the individual structure functions of kplr008703536, kplr011021406, and kplr012208602. A visual inspection of the corresponding light curves in Fig. \ref{fig:AllLightcurves} suggests that these objects do not exhibit significant levels of variability. This observation is borne out by the structure functions of these objects, which are mostly flat and featureless for the duration of observed time-lags. On long timescales ($\sim 100$ d) the structure function begins to exhibit low amplitude `wiggles'. It is well known from the theory of ACF estimation \citep{bro02} that similar wiggles occur on timescales comparable to the length of the time-series and are not significant.

\section[]{Comparison}\label{Discussion}

We observe that not all AGN exhibit DPL parameter $\gamma \sim 2$, consistent with an AR(1) process. Similar observations have been reported by others. In this section we compare our results with those from other AGN variability studies performed using both ground- as well as as space-based instruments. We begin by comparing our results with other studies of \textit{Kepler} AGN.

\subsection[]{\textit{Kepler} AGN Variability Studies}

We have shown that the AGN in this study exhibit PSD with shapes that are superficially similar to that expected from a simple variability model, such as the damped random walk or AR(1) process---a flat power spectrum on long timescales that turns into a $1/f^{\gamma}$ power-law section after frequencies higher than $1/\tau$. However while the AR(1) requires the value of $\gamma$ to be exactly $2$, we find a wide range of $\gamma$ values inconsistent with the AR(1) model. Both \cite{car12} and \cite{mus11} obtained values of the PSD slope for the Seyfert I AGN Zw 229-15 or kplr006932990. \cite{car12} applied various PSD models and found that the best fit estimate of the PSD slope is $\gamma \sim 2.8$ for both their `knee model' as well as their `broken power-law' model. Their estimates of the break timescale ranged from $\tau = 92 \pm 27/21$ d for the `knee' model to $\tau = 43 \pm 13/10$ d for their `broken power-law' model, which is in good agreement with our estimate of $\gamma = 2.7$ while we find a somewhat shorter characteristic timescale $\tau = 27.5$ d. \cite{mus11} computed the PSD slope for 4 individual quarters and found an average PSD slope of $\gamma = 3.115$, significantly steeper than our estimate. It should be noted that these results have been obtained using different versions of the final \textit{Kepler} data product. \cite{car12} were able to use ground-based photometry from the reverberation mapping campaign of \cite{bar11} along with a customized aperture to stitch together the 4 quarters of light curve data used in the study. \cite{mus11} calculate PSD fits for 4 individual quarters of \textit{Kepler} data but use the un-processed SAP light curve in their estimate of the PSD. We used PDCSAP fluxes (Data Release 21+) from all 14 available quarters of data in our analysis. These fluxes, calibrated using the newest version of the PDC module, should suffer only marginally from instrumentation issues. Thus, it is likely that the range of PSD estimates observed is a result of the subtle but important differences between the data products used. Given that all 3 studies have used different analysis techniques but arrive at similarly steep PSD slope estimates, it is clear that the PSD slope is robustly steeper than $1/f^{2}$ and is irreconcilable with the AR(1) process. Most recently, \cite{kel14} attempt to use the Kalman filter to apply a general ARMA model to a single quarter of data from kplr006932990 and find that the observed variability is best-fit by an ARMA process with 6 Auto-Regressive and 4 Moving Average components---a much more complicated stochastic process than a simple damped random walk.

\cite{mus11} also obtained PSD slope estimates for 3 other AGN: kplr012158940, kplr011178007, and kplr002694186. Their analysis was performed on the SAP fluxes observed during individual quarters. The averages of their estimates of the PSD slopes: $\gamma = 2.67$ (kplr012158940) and $\gamma = 2.92$ (kplr011178007) agree fairly well with our estimates in Table \ref{tab:Results}; however, they observe a much steeper $\gamma = 2.845$ for kplr002694186 than our estimate of $\gamma = 2.1$. Both kplr012158940 and kplr011178007 exhibit strong variability--most of the `signal' observed in the uncalibrated SAP light curves for both objects was intrinsic to the source and retained in the calibrated PCDSAP light curves. On the other hand, the calibrated PDCSAP light curve of kplr002694186 lacks the large amplitude flux variations seen in the original SAP light curve, indicating that those variations were instrumental effects that may have inadvertently introduced power at lower frequencies in the PSD computed by \cite{mus11}, artificially making the PSD slope steeper than that computed in our analysis.

\cite{ede13} have examined the variability properties of the light curve of the blazar kplr006690887. Using segments of data spanning $\sim 5.5$ d, \cite{ede13} found that while power law PSDs yield unacceptable fits for the observed PSD, the least unacceptable model is a bending power law PSD with slopes $\gamma = 1.30 \pm 0.14$ and $2.86 \pm 0.34$ if the highest frequencies are excluded from the analysis. Using the full light curve, we find that our fit is of very poor quality with $\gamma = 2.2$. However, flaring is not an AR(1) process. An AR(1) process is completely incapable of producing the flare features observed in this AGN. Ground based studies of blazar variability, such as \cite{rua12}, have attempted to model blazar variability with an AR(1) process using 101 blazar light curves drawn from the LINEAR near-Earth asteroid survey. The LINEAR survey collected time series data over a period of 5.5 yr with two 1.01 meter telescopes giving a r-band survey depth of 18 mag at 5 $\sigma$. Sources in the survey within $10^{o}$ of ecliptic have 460 observations on average, while sources further from the ecliptic only have 200 observations over the duration of the survey, although the cadence can occasionally be as high as once every 15 min. While \cite{rua12} find that the blazar light curves observed by them are well fit by the AR(1) process with decorrelation timescales ranging from $< 1$ d to $\sim$ 1000 d with a peak at $\sim$ 100 d, we caution that the LINEAR survey suffers from both highly irregular, sparse sampling patterns, and large photometric uncertainties.

\cite{weh13} and \cite{rev14} have estimated the PSD slopes of the radio-loud AGN kplr011021406, kplr011606854, kplr01228602, and kplr010663134 in Table \ref{tab:Results}. \cite{weh13} perform a PSD analysis of the un-corrected SAP flux light curve of each object on a per quarter basis. They also supply the same analysis for `corrected' and `end-matched' versions of the light curve for each quarter, where they corrected the SAP data by removing some of the instrumental effects and windowed the data by removing a linear trend from the data to make the first and last point of each quarter have identical flux value. In the case of kplr011021406, they concluded that this Seyfert 1.5/FSRQ is variable and derive a mean PSD slope $\gamma = 1.8$ with standard deviation calculated from each quarter equalling $0.2$. Individual estimates of their PSD slopes for each quarter range from $1.8$ on the low side to $2.0$ on the higher end. In comparison, using the PDCSAP light curve from Data Release 21+, we find no intrinsic variability in the light curve of this object suggesting that it is essential to use the PDCSAP flux when studying light curve variations in order to properly remove spacecraft induced trends from the data. For the FSRQ kplr011606854, they derive a mean PSD slope of $\gamma = 2.0$, which is significantly flatter than the value estimated by us ($\gamma = 3.0$). The remaining AGN, kplr010663134, has slope $\gamma = 1.9$ which in good agreement with our result in Table \ref{tab:Results}. Subsequently \cite{rev14} used a stitching algorithm in their analysis of the multi-quarter light curves of these AGN. After stitching together 11 quarters of SAP flux data, \cite{rev14} obtain PSD slope estimates of $\gamma = 1.9$ (kplr011606854) and $\gamma = 1.7$ (kplr010663134)--significantly different from the values in Table \ref{tab:Results}. This discrepancy is partially caused by the use of the SAP fluxes by \cite{rev14} as opposed to PDCSAP fluxes by us. More importantly, the DPL model is just unable to fit the observed structure functions of AGN that have pronounced flares in the light curve.

In conclusion, we find that our results are in good agreement with the previous studies of \cite{mus11,ede13,weh13,rev14} with most differences being attributable to the variations in data-length and data-versions between the light curves used by us and previous efforts.
 
\subsection[]{Ground-Based Studies that Use the AR(1) Process}

Previous variability studies using ground-based data suggest that the AR(1) process is adequate to characterize AGN variability given the measurement uncertainties and sampling pattern typical of ground-based studies. The original estimation performed by \cite{kel09} using the standard state-space representation of the AR(1) process \citep{bro02} indicated that the R- and B-band optical variability observed in their sample of 109 quasars and Seyferts was well modeled by AR(1) processes with a range of decorrelation timescales strongly indicative of a connection between thermal fluctuations occurring on the AGN accretion disk and the observed variability. The 109 AGN used in this study include 59 MACHO quasars observed with sampling rates of once every 2--10 d over 7.5 yr, 42 PG quasars observed once every 40 d over 7 yr, and 8 Seyferts observed once every 1--10 d resulting in their data set having no cadence higher than $\sim$ 1 d$^{-1}$. The best-fit characteristic timescales for their objects range from $\sim$5--20000 d with a median value of 540 d. It should be noted that while the shortest interval between 2 flux measurements in the dataset may be $\sim$1 d, as is the case in all ground-based studies, this interval is not constant, i.e. the median sampling interval is substantially higher than $\sim 1$ d. Looking at the \textit{Kepler} structure functions in Fig. \ref{fig:StructureFunctions}, this dataset and other ground-based datasets do not sample the structure function very well below $\sim 10$ d making these studies less sensitive to the short timescale ($\lesssim 5$ d) properties of AGN light curves.

\cite{koz10} used the Press-Rybicki-Hewitt maximum likelihood technique \citep{ryb92} to model a much larger sample of $\sim$ 88700 I- and V-band candidate quasar light curves from the OGLE-II and -III surveys as a set of multivariate Gaussian distributions parametrized by correlation matrices with the structure expected from the AR(1) process. The OGLE surveys that the data were obtained from collectively span about 12 yr and offer between 100 (V-band) to 750 epochs per light curve, giving an average sampling rate of once every 6 d in the I-band. By using a set of known quasars \cite{koz10} were able to suggest a parameter-space cut to select quasars. The characteristic timescales found by them range from $\sim$10--3000 d and are in agreement with the range observed by \cite{kel09}.

Similar results were obtained using data from the SDSS Stripe 82 data set. The SDSS Stripe 82 data set spans about 10 yr and provides about 65 epochs of data for each object in the survey giving a sampling rate of about 6--10 observations every year. When comparing results from the SDSS survey to our work, it should be noted that the typical SDSS quasar is much more luminous ($M_{i} \lesssim -24$) than the AGN observed by \textit{Kepler}. \cite{mac10} used the PRH algorithm to model the light curves of the $\sim 9000$ quasars in the SDSS Stripe 82 time domain survey as AR(1) processes and observed the same range of characteristic timescales as \cite{kel09} and \cite{koz10}; most quasars have decorrelation timescales in the range of $\sim 3-3000$ d. These results were put to the test by combining data from the POSS survey ($\sim 50$ yr baseline with a handful of points per light curve) with the Stripe 82 data by \cite{mac12}. These authors found that the AR(1) process continued to fit the observed light curves well and yielded decorrelation timescales in the range of $5-2000$ d for the light curves. Based on these results, \cite{mac11}, \cite{but11}, and \cite{cho14} have proposed that quasars may be selected from variable point sources, such as non-periodic variable stars, using the variability parameters as a selection criteria. \cite{kel13} have shown how variability may be used as an estimator of the black-hole mass of the AGN. The AR(1) process has even been successfully applied by \cite{sob14} to $\gamma$-ray AGN variability observed by the \textit{Fermi} space telescope.  

Recently \cite{zu13}, \cite{and13}, and \cite{gra14} have rigorously tested the AR(1) process on the timescales probed by the surveys used in the previous studies. Using the PRH formalism adopted by \cite{koz10}, but using more generalized forms for the structure of the covariance matrix employed, \cite{zu13} concluded that while the OGLE light curves used in the study are consistent with the AR(1) process on the longest timescales probed by the OGLE data ($\sim 7$ yr), on short timescales (less than a few months) there is some evidence for a steeper than $1/f^{2}$ PSD. Furthermore, they concluded that while the AR(1) process produces acceptable fits to the light curves, on the whole the simple models used in that study are not well constrained by the light curves. \cite{zu13} concede that the quality of the photometry may be responsible for the mixed results obtained by them when fitting the various covariance matrix models, i.e the scatter observed by them in the slope parameter fit may be either intrinsic to the light curves and evidence for varied behavior between the objects, or a consequence of errors in the determination of the photometric uncertainties. A comprehensive and systematic Bayesian study of SDSS Stripe 82 quasar light curves performed by \cite{and13} concluded that more sophisticated models, such as AR(2), GARCH(1), and ARMA(1,1), are not required to model light curves. Of the 6304 quasars examined in this study, an overwhelming majority--5047 light curves--were best modeled by an AR(1) process while only a handful required a more sophisticated approach. 

While a large number of studies have indicated that the AR(1) process is sufficient to model AGN variability, these studies have used comparatively sparsely sampled light curves ($10-100$ points over several years) that do not well sample the light curve on short timescales ($\Delta t \lesssim 5$ d). There are indications that more complex stochastic models are required to characterize variability on short timescales. \cite{sch12} use the simple power-law form of the structure function given in Eq. \eqref{eq:SchmidtSF}
to model the structure function of Stripe 82 quasars. They compute the structure function using
\begin{equation}\label{eq:SchmidtSFCalc}
SF(\Delta t_{i,j}) = \langle \sqrt{\frac{\pi}{2}} |\Delta m_{i,j}| -\sqrt{\sigma_{i}^{2}+\sigma_{j}^{2}} \rangle_{\Delta t},
\end{equation}
where $\Delta m$ is the magnitude difference over the time interval $\Delta t$. While our structure functions is \textit{quadratic} in flux difference, the structure function of \cite{sch12} is \textit{linear} in magnitude difference. Our correlation strength parameter $\gamma$ appears in the light curve PSD model, while the correlation strength parameter $\lambda$ of \cite{sch12} appears in the structure function model making it impossible to directly compare our $\gamma$ with their $\lambda$. However, \cite{sch12} find that Stripe 82 quasars are fit by $\lambda$ values ranging from 0.0 to 1.2 with a peak at $\lambda = 0.25$. The range of $\lambda$ values found by \cite{sch12} suggests that a simple stochastic process model like the AR(1) process is unlikely to work well for all objects. \cite{gra14} use a wavelet analysis technique known as the Slepian Wavelet Variance (SWV) to study 18000 quasars from the Catalina Real-time Transient Survey (CRTS) and SDSS Stripe 82. By comparing the expected Slepian Wavelet Variance from an AR(1) process to that observed for real quasars in S82 and CRTS, \cite{gra14} find that there is a clear indication that the AR(1) process fails to characterize AGN variability on short timescales. 

While one cannot ignore the success of the AR(1) process at modeling the time variability behavior of AGN on the timescales probed by MACHO, OGLE, SDSS, etc., we caution that it may be impossible to observe deviations from the AR(1) process without probing the short timescales available through \textit{Kepler}.

\section[]{Conclusions}\label{Conclusions}

Individual \textit{Kepler} AGN light curves show a wide range of behavior that can be loosely grouped into 5 categories: stochastic-looking, somewhat stochastic-looking+weak oscillatory features, oscillatory features dominant, flare features dominant, and not-variable. Some light curves appear to transition from one variability state to another, suggesting that AGN light curves may not be stationary in the statistical sense (Sec. \ref{StitchResults} and Fig. \ref{fig:AllLightcurves}). We estimate PSD slopes ranging from $\gamma = 1.7$ to $\gamma = 3.1$ with $3$-$8$ objects having slopes close to that of the AR(1) process ($\gamma \equiv 2.0$). Seven of the remaining objects have slopes significantly steeper than $\gamma = 2.5$ suggesting strongly correlated non-AR(1) light curves, while two have $\gamma < 1.9$. Four out of 5 of the AGN that exhibit oscillatory features in the light curve have PSD slope close to that of the AR(1) process, while the AGN that exhibit flares in the light curve are poorly fit by the DPL model (Sec. \ref{Results} and Table \ref{tab:Results}). 

A broad dip feature can be observed in the structure function on timescales ranging from $\sim$10--100 d in the rest frame of 12 of the 20 AGN. This dip feature in the structure function corresponds to increased correlation on these timescales in the light curve, i.e. flux measurements on timescales ranging from $\sim$10--100 d are closer than they should be. Due to the wide range of redshifts of these objects, this timescale range is not identical in observed frame of these objects, suggesting that this feature is probably not caused by some sort of instrumentation issue (see Fig. \ref{fig:kplr006932990-SF1}). Some AGN exhibit varying structure function slopes, implying that the value of $\gamma$ in the DPL model varies on different timescales (see Fig. \ref{fig:kplr009215110-SF1}). Lastly, not all AGN show variability at the levels probed by \textit{Kepler}, confirming previous findings that flux variability is seen in most but not all AGN (see Fig. \ref{fig:StructureFunctions}). 

There is no clear relationship between the PSD slope and object type or redshift, although the sample size is too small to draw any definitive conclusion. We conclude that while enough of our light curves are consistent with an AR(1) process for the damped random walk model to be an appealing choice--especially for poorly sampled light curves--there are enough interesting features present in the light curves to warrant a more detailed analysis.  

The AGN light curves seen in \textit{Kepler} suggest that AGN variability is a very complex phenomenon with individual light curves looking very different. We will perform a recalibration of the \textit{Kepler} AGN light curves to determine which light curve visual features are persistant. We will apply more sophisticated statistical models drawn from the field of time series analysis such as the ARMA process model \cite{ham94}. The AR(1) process or DRW model can determine `if' an AGN is varying, but is not helpful in determining `why' the AGN is varying.

\section*{Acknowledgments}

We acknowledge support from NASA grant NNX14AL56G. We thank Coleman Krawczyk for his help with data visualization and Dr. N.P. Ross for his insightful comments on this paper.

This paper includes data collected by the Kepler mission. Funding for the Kepler mission is provided by the NASA Science Mission directorate. All of the data presented in this paper were obtained from the Mikulski Archive for Space Telescopes (MAST). STScI is operated by the Association of Universities for Research in Astronomy, Inc., under NASA contract NAS5-26555. Support for MAST for non-HST data is provided by the NASA Office of Space Science via grant NNX13AC07G and by other grants and contracts.

\label{lastpage}

\end{document}